# Local Analysis of Dissipative Dynamical Systems

**Abstract**

Linear transformation techniques such as singular value decomposition (SVD) have been used widely to gain insight into the qualitative dynamics of data generated by dynamical systems. There have been several reports in the past that had pointed out the susceptibility of linear transformation approaches in the presence of nonlinear correlations. In this tutorial review, the local dispersion along with the surrogate testing is suggested to discriminate nonlinear correlations arising in deterministic and non-deterministic settings.

*Author*

Radhakrishnan Nagarajan
Center on Aging
University of Arkansas for Medical Science
629 S. Elm Street, Room No: 3105
Little Rock, AR 72205
Email: nagarajanradhakrish@uams.edu

## 1. Introduction

A nonlinear deterministic dynamical system can be represented by the differential equation,

$$\frac{dx}{dt} = G(x, \boldsymbol{a}) \tag{1}$$

where $x = (x_1 x_2 \ldots x_D) \in U$, represent the states of the system in state space $U$. The vector field $G$, maps $U$ into the tangent space $TU$, represented by $G : U \rightarrow TU$. The evolution of the system with respect to time $t$ is represented by the flow $x_t = \boldsymbol{j}(x_0)$, for the initial condition $x_0$.

For dissipative dynamical systems, the flow settles down on an attractor, in a $C^k$ *manifold* of dimension $m$ in $M$. $M$ is *locally* Euclidean $R^m$, and can be differentiated at most $k$ times hence $C^k$. The dimension of the manifold $M$ is generally lesser than that of the original state space $U$, and is governed by the strength of the nonlinear interactions between the state variables and the system parameter $\boldsymbol{a}$, in (1). There have been several instances where experimental systems consisting of several state variables converge onto low-dimensional manifolds after an initial transient behavior [Alligood et al, 1996]. The method of delays [Takens, 1981; Mane, 1981, Sauer et al, 1991, Packard, 1980] provides a way to reconstruct a state space (phase space) in $R^{2m+1}$ under certain *generic* conditions by sampling a single dynamical variable. The reconstructed phase space is topologically equivalent to the original state space. Invariants and geometry of the original state space is preserved in the reconstructed phase space. A rigorous mathematical treatment and extensions of the embedding procedure can be found elsewhere [Takens, 1981; Noakes, 1991, Stark et al, 1997].

The one-dimensional time series data sampled from a dynamical system can be represented as $\{w_n\}$, $i=1...\text{N}$, $w_n = w(\boldsymbol{j}_n(x)) \in R$, where $w$ is the measurement function. Given the limited number of data points (N), a proper choice of embedding dimension ($d$) and time delay ($\boldsymbol{t}$) is essential for attractor reconstruction [Kennel et al, 1992; Liebert & Schuster, 1988; Frasier and Swinney, 1986; Cellucci et al, 2003]. Such a choice prevents self-intersection of the trajectories in the phase space and ensures proper unfolding of the attractor. The method of false-nearest neighbors (FNN) and the auto-correlation function had been used extensively to determine the embedding dimension $d$, and the time delay $\boldsymbol{t}$, respectively. However, convergence of techniques such as FNN is sensitive to the number of samples and the noise in the given data. The embedding dimension had also been observed to increase with external noise [Hegger et al, 1999]. The auto-correlation is a measure of linear dependencies and not sensitive to nonlinear dependencies. The mutual information had been suggested as an alternative to overcome these shortcomings [Frasier and Swinney, 1986]. However, the computation of the mutual information is non-trivial. As a remark, it should be noted that the linear correlation (L) is related to the mutual information (M) by the expression $M = -0.5 \times \log(1 - L^2)$ in the case of normally distributed process. The error term in the case of linearly correlated process is assumed to be normal and the second-order statistics is sufficient statistics in estimating its optimal parameters, given by the Yule-Walker equations [Proakis and Manolakis, 1995]. However, this is not true when the distribution of the error term is non-normal, which is the case in nonlinear processes [Tong, 1990; Theiler et al., 1993].

Recent reports have argued that the choice of the embedding dimension is based on the problem at hand [Grassberger et al, 1991; Kantz and Schreiber, 1997]. Cellucci et al, 2003 have reported that a combination of mutual information and false-nearest neighbors can be useful in determining the optimal embedding parameters. In this report, we implicitly assume a proper choice of the embedding parameters, and hence shall not address this issue in the subsequent discussions.

The vectors reconstructed using the embedding parameters ($d$, $t$) can be represented as the rows of the $N - (d-1)t$ x $d$ *trajectory matrix* ($\Gamma$) with elements $g_{kj} = w_{k+(j-1)t}$, given by

$$\Gamma = \begin{bmatrix} g_{11} & . & . & g_{1d} \\ . & . & . & . \\ . & . & . & . \\ g_{N-(d-1)t1} & . & . & g_{N-(d-1)td} \end{bmatrix} = \begin{bmatrix} v_1 \\ . \\ . \\ v_{N-(d-1)t} \end{bmatrix} \qquad (2)$$

where $v_k = (w_k, w_{k+t}, ..., w_{k+(d-1)t})$, $1 \le k \le N - (d-1)t$

The $i^{th}$ column of the trajectory matrix represents the samples along the $i^{th}$ dimension, $i = 1...d$. Alternatively, each row can be represented by a point in the phase space.

The *singular value decomposition* (SVD) had been proposed in the past for extracting the qualitative dynamics from the trajectory matrix $\Gamma$ [Broomhead and King, 1986]. Broomhead and King (BK) proposed an approach to capture the dynamics of the trajectories in the phase space by identifying the dominant eigen-values of the covariance matrix. BK's approach assumed that the embedded manifold is restricted to a subspace of $R^d$, where $d$ represents the rank of the covariance matrix, (Section 2). In other words, the rank of the covariance matrix was assumed to form an upper bound to the subspace explored by the deterministic part of the trajectory. Since external noise is unavoidable in experimental data the eigen-values were believed to drop to a *noise floor*. A plot of the sorted, normalized eigen-values with respect to the embedding dimension was found to exhibit a sharp transition in the case of data obtained from nonlinear deterministic dynamical systems. The embedding dimension at which this transition occurred was chosen as the optimal embedding dimension. Mees et al. 1987, demonstrated the shortcomings of this technique, when applied to data sets generated by nonlinear deterministic processes. Subsequent reports also discussed the inherent assumptions and limitations of SVD when applied to chaotic data sets [Palus and Dvorak, 1992]. In the present study, the close relation between the eigen-decomposition of the trajectory matrix and the spectral power is elucidated. A partitioning approach is suggested to minimize the effect of nonlinearity [Nagarajan, 2003]. The surrogate testing is used along with the partitioning technique to discriminate nonlinear correlations rising from nonlinear deterministic and non-deterministic settings. The description of the data sets to be discussed is enclosed under Appendix, F. The number of points in the data sets were fixed at (N = 2048) to prevent any discrepancies that can be introduced by phase-randomization, which is an inherent part of the surrogate generation algorithms.

The report is organized as follows: in Section 2, the close relation between the SVD of the trajectory matrix and the power spectral estimation of linearly correlated process is discussed. The fact that SVD of the trajectory matrix reflects the spectral content as opposed to the dimensionality of the underlying dynamical system is elucidated with several examples, including those published already [Mees et al., 1987]. It is argued that a possible explanation for the failure to observe decay in the magnitude of the eigen-values in the case of nonlinear deterministic chaotic processes might be due to its characteristic broad-band power spectrum. SVD being a linear transform is susceptible to nonlinear correlations. To minimize the nonlinear effects a clustering technique is proposed in Section 4. However, it should be noted that nonlinear correlations can arise in deterministic as well as non-deterministic settings. The method of surrogates, Section 3, is used along with the clustering technique to discriminate nonlinear correlations arising from diverse settings. The phase space is partitioned into non-overlapping clusters. The local dispersion, given by the product of the eigen-values of the local trajectory matrix, is estimated for each of the clusters, Section 4. Two different techniques are proposed for statistical discrimination, Section 3. In the first, the statistical significance is determined by comparing the distribution of the *local dispersions* obtained on the empirical (original) data directly compared to those obtained on a surrogate realization. In the second approach the *total dispersion* obtained on the original data is compared to those obtained on an ensemble of surrogate realizations. Parametric, non-parametric and resampling statistics are used to estimate statistical significance. Issues such as determining the required number of clusters and the effect of eigen-values comprising the noise floor are also discussed in the same.

## 2. Singular Value Decomposition of the trajectory matrix **G**

Singular value decomposition is a linear transformation and is used to decompose the given matrix $\Gamma_{pxq}, p > q$, into $\Gamma = U\Sigma V^{T}$, where $U_{pxq}$ and $V_{qxq}$ are the left and right orthogonal matrix and $\Sigma_{qxq}$, the diagonal matrix [Golub and van Loan, 1996]. The diagonal elements of $\Sigma_{qxq}$ are the desired *singular values*, also known as *eigen-values*, represented by $\lambda_{i}, i = 1...q$ with $\lambda_{i} \geq \lambda_{i+1}$. The matrices $V$, $\Sigma$ and $U$ correspond to rotation, stretching and rotation in that order. The SVD of $\Gamma$ is related to the eigen-decomposition of the symmetric matrices $\Gamma^{T}\Gamma$ and $\Gamma\Gamma^{T}$, as $\Gamma^{T}\Gamma v_{i} = \lambda_{i}^{2} v_{i}$ and $\Gamma\Gamma^{T} u_{i} = \lambda_{i}^{2} u_{i}$. The non-zero eigen-values of $\Gamma^{T}\Gamma$ are the same as that of $\Gamma\Gamma^{T}$, and determine the rank of $\Gamma$. The singular values of $\Gamma$ are the square roots of the eigen-values of $\Gamma^{T}\Gamma$. Symmetric matrices by definition are positive semi-definite and have non-negative eigen-values, i.e. $\lambda_{i} \geq 0, \forall i$. However, in the presence of numerical and experimental noise $\lambda_{i} > 0, \forall i$, Section 2.2.

### 2.1 SVD of **G** and Spectral Estimation

Power spectral techniques have proven to be powerful tools in identifying the dominant frequency components in linearly correlated processes. The close connection between the power spectrum and the auto-correlation function is given by the Wiener-Khintchine theorem (Appendix, B). Power spectral estimation can be broadly classified into parametric, non-parametric and subspace methods [Proakias and Manolakis, 1995]. The objective in this section is to establish the relationship the between spectral estimation and eigen-decomposition of the trajectory matrix and not to critically evaluate the spectral estimation techniques. Analysis of linearly correlated stationary processes begins with the Wold decomposition, which states that any stationary process can be represented as the sum of a deterministic and non-deterministic part (*auto-regressive moving average*, ARMA process). The AR (auto-regressive) and the moving-average (MA) represent the deterministic and non-deterministic parts respectively. For example, a real sinusoidal signal corrupted with white noise can be modeled as a ARMA $(p,p)$ processes, with identical AR$(p)$ and MA$(p)$ parameters, (Appendix, D). Subspace decomposition methods such as *Pisarenko Harmonic Decomposition* (PHD) [Pisarenko, 1973], estimate the pseudospectrum of harmonically related sinusoids corrupted with zero mean additive white noise, by eigen-decomposition of the auto-correlation matrix. Techniques related to the PHD are *Multiple SIgnal Classification* (MUSIC) and *Estimation of Signal Parameters via Rotational Invariance Technique* (ESPRIT) [Proakis and Manolakis, 1995]. The mathematical framework behind PHD relies on Caratheodory's *uniqueness result* (Appendix, C) [Caratheodory and Fejer, 1911; Sidiropoulos, 2001], which provides a bound on the size of the toeplitz matrix, necessary to extract the $p$ distinct frequency components. Consider a toeplitz matrix of size $m$, with $m > p$, its eigen-decomposition yields $p$ dominant eigen-values and their corresponding eigen-vectors span the *signal subspace*. The *noise subspace* is spanned by the remaining $(m\text{-}p)$ eigen-vectors. It might be interesting to note that the subspace decomposition can be achieved by the SVD of the trajectory matrix [3] with the embedding dimension chosen as $(d = m > p)$ and time delay $(t = 1)$, (Appendix, D). In BK's report, it was proposed that the SVD of the trajectory matrix could be used to obtain an upper bound on the embedding dimension of the dynamical system generating the given data. In the following discussion we shall illustrate that the SVD of the trajectory matrix reflects the spectral content as opposed to the dimensionality of the dynamical system generating it. Several examples including those published by Mees et al. 1987 are included.

### 2.1.1 Deterministic Data

*Periodic Data*

The logistic and henon maps (Appendix, F) with parameters $r = 3.55$ and ($a = 1.03$, $b = 0.3$) respectively, were used to generate period eight limit cycle ($p = 8$, N = 2048, $t = 1$). While (F1) and (F2) are nonlinear dynamical systems, (F3) represents a linear combination of harmonically related sinusoids (Appendix, F). Periodic data sets by definition are one-dimensional and their frequencies are harmonically related (Appendix, D), i.e. each frequency component can be expressed as the product of the fundamental and a rational number (winding number), such frequencies are said to be *commensurate*. The trajectory matrix, $\Gamma$ was constructed by varying the embedding dimension $d = 4$, 8, 12,16 with fixed $t = 1$, for each of the data sets. The variation of the sorted normalized eigen-values $l_i^* = l_i / \sum_{i=1}^{d} l_i$ obtained by SVD of $(\sqrt{N - (d-1)t})^{-1} \Gamma$ with respect to the normalized index $(d_i / \max(d_i)$ [Mees et al. 1987], for the three data sets shown in Fig. 1a, 1b and 1c, respectively. A plot of $l_i^*$ reveals no significant decrease in the magnitude of the eigen-values is observed for the $d = 4$, 8, whereas a marked drop is observed for $d = 12$, 16 for the three data sets. More importantly, a sharp transition occurs between $d = 8$ and d = 9, for the three data sets agreeing with Caratheodory's uniqueness result (Appendix, C). While the SVD of $\Gamma$ failed to identify the dimensionality of the dynamical system in the above cases, it does reflect the periodicity of the data, hence its spectral content. Period doubling is a widely observed phenomenon and had been proposed as a possible route to chaos [Feigenbaum, 1979]. Each period doubling is accompanied by appearance of subharmonic peaks in the power spectrum. In the asymptotic limit, one observes a continuum of frequencies (broad-band power spectrum) characteristic of white noise. For a periodic limit cycle of period $p = 2^n$ a sharp transition in the magnitude of $\boldsymbol{I}_i^*$ is expected to occur between $d = 2^n$ and $d = 2^n+1$. However, it should be noted that with increasing period doubling, the required embedding dimension $(d)$ to determine the periodicity increases *exponentially*.

### b. Quasi-periodic Data

Unlike periodic data sets, the frequencies of quasiperiodic data are not harmonically related. An example would be the superposition of sine waves with *incommensurate* frequencies (F4), Fig. 1d. The ratio of the frequencies in this case is an irrational number (winding number). Quasiperiodic behavior can also be exhibited by nonlinear dynamical systems [Alligood et al, 1996]. The phase space representation resembles a *torus*. The Caratheodory's uniqueness result does not impose constraint that the frequencies need to be harmonically related. Therefore, the $p$ distinct frequency components can be extracted by embedding the one-dimensional signal with $(d = m,\ m > p)$ and $\boldsymbol{t} = 1$. As in the case of periodic data, SVD of the trajectory matrix can be used to determine the dominant frequency components and not the dimensionality of the phase space. A plot of $\boldsymbol{I}_i^*$ reveals no significant decrease in the magnitude of the eigen-values is observed for the $d$

= 4, 8, whereas a marked drop is observed for $d$ = 12, 16. More importantly, a sharp transition occurs between $d$ = 8 and d = 9, Fig.. 1d.

As a remark, it should be mentioned that periodic (a) and quasiperiodic (b) behavior can be generated by linear and nonlinear processes. Thus, no conclusion can be drawn about the nature of the process from the observed data sets.

*c. Chaotic Data:*

Unlike periodic and quasiperiodic data, which show prominent peaks in their power spectrum, chaotic data sets exhibit broadband spectra characteristic of random noise. Consider the logistic map (F1) $r$ =3.58, which corresponds to a *noisy* period four limit cycle. The power spectrum (Fig. 2b) consists of a broad-band interspersed between prominent peaks characteristic of period four limit cycle, $r$ = 3.5, Fig. 2a. The plot of $l_i^*$, reveals a similar trend for $r$ = 3.5 and $r$ = 3.58, for the first four points ($d$=1…4) corresponding to the prominent peaks in their power spectrum (period 4), however the trends change significantly for $d > 4$, Fig. 2b. In the case of $r$ = 3.58, the plateau is composed of harmonics that correspond to frequencies which form the broad band (noisy part) of the power spectrum. Thus one fails to observe any sharp transition in the magnitude of the eigen-values. A chaotic time series (N = 2048) points was generated for the logistic equation (F1) [Mees et al, 1987] and henon map (F2) with ($r$ = 4.0, N = 2048, $t$ = 1) and ($a$ = 1.4, $b$ = 0.3, N = 2048, $t$ = 1) respectively. The normalized eigen-values $l_i^*$ for the various choices of the embedding dimension $d$ = 4, 8, 12, 16, 24, and 30 is shown in Fig. 3a and Fig. 3c respectively. A plot of the sorted $l_i^*$ failed to exhibit a sharp transition with increasing $d$ for the chaotic data sets. The iterated amplitude adjusted Fourier transform (Section 3.1.3) [Schreiber and Schmitz, 1996], was used to generate constrained randomized versions of the data sets. While the power spectrum was retained, the temporal structure and hence the dynamics was completely destroyed in the surrogate realization. It is important to note that unlike the chaotic data sets, their surrogate realizations are not finite dimensional. The normalized eigen-values $l_i^*$ for the various choices of the embedding dimension $d = 4, 8, 12, 16, 24,$ and 30, for the IAAFT surrogate is shown in Fig. 3c and Fig. 3d respectively. From Figs. 3a, 3b, 3c and 3d it is evident that the SVD of $\Gamma$ reflects the spectral content and not the dimensionality of the dynamical system.

### 2.1.2 Non-Deterministic Data

*a. Narrow band Auto-regressive moving average (ARMA) process*

A sinusoidal signal corrupted with additive noise (F5) can be modeled as an auto-regressive moving average process (ARMA) with identical AR (auto-regressive) and MA (moving average) parameters (Appendix, D2). The parameters of the ARMA process (F4) were chosen so as to represent a sinusoid of period four corrupted with additive white gaussian noise with zero mean and variance ($a = 0.7$, F5) The power spectrum of the narrow-band ARMA process and that of its IAAFT surrogate is shown in Fig. 4a. The normalized eigen-values $l_i^*$ for the various choices of the embedding dimension $d = 2, 3, 4, 5$ and 6, with time delay $t = 1$, (N = 2048) is shown in Fig. 4b. The susceptibility of the surrogate techniques to narrow-band processes with long coherence time had been reported earlier [Theiler et al., 1993]. The discrepancy in the power spectrum between the given data and its IAAFT surrogate (Sec. 3.1) is reflected by the plot of the normalized eigen-values. However, reconstructing the surrogate retaining the dominant eigen-values ($d = 1...4$) in this case can minimize the discrepancy in the power spectrum.

*b. Nonlinear transform of correlated noise*

For nonlinearly correlated noise (F7) described in [Schreiber & Schmitz, 1996], one fails to observe a prominent decrease in the relative magnitude of the eigen-values with increasing dimension, Fig. 3e. Unlike the narrow band ARMA, the power spectrum of the nonlinearly correlated noise is broad-band similar to that of its deterministic chaotic counterpart. As expected, the spectral content and hence the magnitude of the normalized eigen-values of the nonlinearly correlated noise (Fig. 3e) and its IAAFT surrogate (Fig. 3f) were similar.

From the above examples, it is evident that the SVD of the trajectory matrix is related to the subspace decomposition of linearly correlated processes and mimics the spectral content in a given data as opposed to its dimensionality. It is inappropriate to apply such linear transformation techniques for the analysis of nonlinearly correlated processes. A possible explanation for the failure to observe a prominent decrease in the magnitude of the eigen-values in the case of chaotic data sets can be attributed to its characteristic broad-band power spectra.

## 2.2. SVD of **G** and phase space geometry

The correlation matrix of $\Gamma$ (Appendix, A) is related to the symmetric matrices $\Gamma^T\Gamma$ and $\Gamma\Gamma^T$, and captures the *linear correlation* of the trajectories in the phase space. The correlation matrix of a $d$-dimensional independent and identically distributed (i.i.d) process is the identity matrix $I_{dxd}$. The identity matrix is a full rank matrix, with $d$ uniformly distributed eigen-values of unit magnitude, represented by a $d$-dimensional sphere. Non-zero value of the off-diagonal elements is indicative of linear correlation among the column vectors and is accompanied by non-uniform distribution of the eigen-values. The eigen-values determine the extent of stretching and contraction along the orthogonal directions. In the analysis of nonlinear systems, the qualitative behavior in the neighborhood of certain class of fixed points is determined by the eigen-decomposition of the Jacobian [Wiggins, 1990]. For a two-dimensional case, it is appropriate to view the correlation matrix as a linear transformation of a unit circle into an ellipse (Appendix, B). Finite size, numerical precision and noise inherent in experimental data render the eigen-values strictly greater than zero, $I_i > 0, \forall i$ and hence the correlation matrix positive definite. As pointed out by BK, in the presence of white noise, the correlation matrix can be represented as an additive sum of the contributions due to the signal and noise, given by $\Gamma_r^T = \Gamma_r^{signal} + \Gamma_r^{noise}$. The random part $\Gamma_r^{noise} = s^2 I$ is a full rank matrix irrespective of the rank of $\Gamma_r^{signal}$. Thus results based on the rank of the correlation matrix are inconclusive. An alternate approach would be to determine the dominant eigen-values of the correlation matrix. The dominant eigen-values are chosen such that they are greater than a pre-defined threshold. The number of dominant eigen-values has been suggested to represent the dimensionality of the dynamical system [Passamante et al., 1989].

However, the choice of the pre-defined threshold is non-trivial and will be discussed in Sec. 4.2.

The sum of the eigen-values of the correlation matrix is given by

$$d = \sum_{i=1}^{d} \lambda_i$$ , where $\lambda_i$, is the eigen-value corresponding to the $i^{th}$ principal

axis of a $d$-dimensional ellipsoid.

The product of the eigen-values of the correlation matrix is given by

$$V = \prod_{i=1}^{d} \lambda_i$$ , where $V$ represents the volume of a $d$-dimensional ellipsoid.

Define $L(V,d) = V + \lambda d$, where $\lambda$ is the Lagrange multiplier. The maximum volume is obtained by partially differentiating the $L(V,d)$, with respect to each of the eigen-values $\lambda_i, 1 \leq i \leq d$ .i.e.

$$\frac{\partial}{\partial \lambda_i} L(V,d) = \frac{\partial}{\partial \lambda_i} (V + \lambda d)$$

Solving the above expression yields a full rank matrix with $\lambda_i = 1, 1 \leq i \leq d$ which corresponds to a $d$-dimensional sphere of volume, $V=1$. Thus a uniform distribution of eigen-values corresponds to a maximum volume and represents an uncorrelated (i.i.d) process. Increasing correlation among the column vectors results in skewed distributions of the eigen-values characteristic of an ellipsoidal shape, $V<1$ (Appendix, A). As a comment, it should be understood that the above discussion implicitly assumes the column vectors of $\Gamma_{N-(d-1)t \, x \, d}$ to be linearly correlated.

*Linear Filtering and phase space distortion*

SVD is often used as a pre-processing step in experimental data analysis, however they can distort the phase space geometry as demonstrated by the following example. As ntoed earlier, SVD of the trajectory matrix (t = 1) is related to spectral decomposition. Therefore, reconstruction using only the dominant eigen-values amounts to linear filtering of the given data. For the logistic map (F1) $r = 3.58$, the plot of the normalized eigen-values with ($d = 40$, $t = 1$, N = 2048) exhibits a transition at $d = 4$, followed by a plateau (Fig. 5a). The eigen-values $l_i$, $1 \leq i \leq 4$, corresponds to the dominant peaks, whereas $l_i$, $5 \leq i \leq 40$ corresponds to the broadband part of the power spectrum, Fig. 5b. Signal reconstruction using the $q$ dominant eigen-values would amount to suppressing the ($d$-$q$) less dominant frequencies and hence linear filtering or de-noising the given data. For the logistic map, retaining the $q = 30$, 20 and 10 eigen-values, explain 98.9%, 96.5% and 92.6% of the variance in the data. The effect of linear filtering on the power-spectrum, is shown in Fig. 5a. It can be seen that the filtering affects the frequencies comprising the broad-band region of the power spectrum. While the power spectrum of the original and the reconstructed signals are similar, there is marked distortion introduced in the phase space representation, Fig. 5b. This is true even when $q = 30$ dominant eigen-values are used for reconstruction. While this example is not meant to criticize the choice of SVD filters, de-noising chaotic data using techniques such as SVD should be regarded with caution.

## 2.2. Nonlinear correlations in diverse settings

Deterministic chaotic data sets obtained from dissipative dynamical systems exhibit stretching, contraction and folding which give rise to manifolds that are usually curved. Manifolds of dissipative dynamical systems by definition are *locally* Euclidean. However, the trajectory matrix is representative of the entire phase space, which does not usually resemble a linear subspace. While nonlinear correlations are an essential ingredient of deterministic chaotic data, they can be observed in deterministic as well as non-deterministic settings. A classic example of the latter as pointed out by Fukunaga and Olsen, 1971 is a nonlinear transform of a random variable. Their report also discussed failure of the linear transformation techniques on determining their intrinsic dimensionality and suggested a local region based approach. Subsequent reports [Passamante et al, 1989] adopted similar approaches. However, the choice of the local regions was more ad-hoc in nature. The emphasis in the current study is on minimizing the nonlinear effects, and not to estimate the intrinsic dimension. While local analysis by partitioning (Sec. 4) can minimize nonlinear effects, it does not distinguish nonlinearities arising from deterministic and non-deterministic settings. Surrogate testing, Sec. 3, is used along with the partitioning technique to overcome this problem. The surrogate testing is related closely to the resampling methods and is discussed in the following section.

## 3. Resampling Methods and Surrogate Realizations

Resampling methods have been used as effective tools to assess statistical significance [Efron, 1982; Efron et al, 1994]. In traditional tests, the statistical measure obtained on a sample, drawn from a given population, is compared to a *theoretical distribution* to assess its significance. However in resampling statistics, one generates an *empirical distribution* by computing the statistical measure on samples generated from the *same* empirical sample, hence the term resampling. Unlike traditional approaches, resampling methods implicitly assumes the given empirical sample to be a representative of the population. This is also regarded as a drawback of this technique. Resampling can be broadly classified into *bootstrap, randomization* and *jackknife* [Efron, 1982; Efron et al, 1994]. In bootstrap, the empirical distribution is generated by random sampling *with replacement*. In randomization the empirical distribution is generated by random sampling *without replacement*. In jackknife, the empirical distribution is generated by leaving a member out each time. The choice of the resampling method is based entirely on the problem at hand. While resampling without replacement is suitable for uncorrelated data, resampling without replacement is appropriate for correlated data. The method of surrogate testing uses resampling without replacement or randomization and has been used successfully to make a statistical inference on the nature of the process generating the given data. A partial list of contributions to surrogate testing includes [Kaplan & Cohen, 1990, Theiler et al, 1992; Rapp et al, 1994; Schreiber & Schmitz, 1996, Kugiumtzis, 2000].

## 3.1 Surrogate Testing

The choice of the measure is governed by the nature of the data. Thus determining the nature of the process generating the given data should be a prior step to choosing a particular measure to quantify it. Parametric statistical measures such as mean and variance can prove to be useful provided the given data is generated by normally distributed uncorrelated process. The choice of non-parametric statistics such as median and kurtosis are useful in the case of data generated by non-normally distributed uncorrelated process. These measures are governed entirely by the distribution and hence inappropriate for the analysis of correlated data. Power spectral techniques can be used to determine the dominant frequencies in data generated by linear processes, Sec. 2. However, it is inappropriate to use power spectral analysis for data sets generated by nonlinear processes, Sec. 2, especially those obtained from nonlinear dynamical system. Invariant measures have been used successfully for the analysis of data generated by nonlinear dynamical processes. However, it has been pointed out that the convergence of invariant estimation algorithms can be observed in diverse settings [Provenzale et al, 1992; Pincus, 1991]. This implies that any conclusion based entirely on their convergence of invariants is incomplete. Recent studies have suggested surrogate testing as an effective tool for statistical discrimination and avoid spurious conclusions [Rapp et al, 1994, Theiler, 1995, Rapp et al, 2001]. The three basic ingredients of surrogate testing are a null hypothesis (H), a discriminant statistic (D) and an algorithm (A) addressing the null (H).

*3.1.1 Random Shuffle Surrogates*

Random shuffle surrogates address the null hypothesis (H) that the given data is generated by uncorrelated random noise. The surrogates are generated by randomization without replacement of the given empirical sample. Therefore, the amplitude distribution of the original data is retained in the surrogate realizations. By retaining the amplitude distribution, we are not restricting ourselves to a random noise with a specified distribution. Thus one can think of the above randomization as a *constrained randomization* [Theiler et al, 1992; Schreiber & Schmitz, 1996], where the constraint is on retaining the amplitude distribution of the empirical data. This should be contrasted against *typical* realization, where the distribution is assumed to be of a particular form. The discriminant statistic (D) is chosen so as to discriminate the empirical data from the surrogate realization in the event of rejecting the null hypothesis. In the present case, a D sensitive to the correlation in the data would reject the null of uncorrelated noise. A significant difference in D between the empirical sample and the surrogate realization implies that one can reject the null hypothesis.

### 3.1.2 Phase-Randomized Surrogates

A more sophisticated null hypothesis would be that the empirical sample is generated by a linearly correlated process. The surrogates addressing this null are generated so as to retain the power spectrum of the empirical sample. Retaining the power spectrum implies retaining the auto-correlation function (Wiener-Khintchine theorem) [Proakis and Manolakis, 1995]. Parameters of a linearly correlated process are estimated from their auto-correlation function (Yule-Walker Equations) [Proakis and Manolakis, 1995]. Therefore, by retaining the power spectrum, the model parameters are retained implicitly. Three different estimators of the auto-correlation function were pointed out [Theiler et al, 1993], namely: unbiased estimator, biased estimator and the circular correlation. Only the circular correlation is preserved in the phase randomized surrogates. However, the three estimators have been found to agree in the asymptotic limit. Unlike typical realizations that assume specific values of the parameters, the constrained realizations are generic, the constraint here, being on retaining the power spectrum of the empirical data. Alternatively, the model parameters of the linearly correlated process are treated as nuisance variables [Schreiber & Schmitz, 1996]. For the case, where the parameters vanish, the empirical data resembles uncorrelated process, given by the noise term. A significant difference between the empirical data and the surrogate realizations, by a chosen discriminant statistic implies that the null can be rejected. The noise term in the linear model is assumed to be normally distributed. The effect of non-Gaussian innovations is discussed elsewhere [Theiler et al, 1993; Tong, 1990]. Nonlinear transforms of linearly correlated noise can result in non-normal distributions and is discussed in the following section.

*3.1.3 Amplitude Adjusted Fourier Transform*

The Amplitude Adjusted Fourier Transform (AAFT) was suggested to accommodate nonlinear transforms of a linearly correlated noise [Theiler et al, 1992]. This addresses a wider variety of process than those discussed before. However, there are constraints on the type of nonlinearity. The nonlinear transform is restricted to a static, monotone nonlinear function. Such nonlinearities are attributed to the measurement device and deemed uninteresting. The objective would be to rule out such trivial nonlinearities. The nonlinear transform is retained in the surrogates by a rank ordering approach and the power spectrum by phase randomization. The discriminant statistic is chosen such that it is sensitive to the temporal structure, hence the dynamical nonlinearities in the data. Recent reports [Schreiber & Schmitz, 1996] have claimed that the AAFT surrogates can introduce flatness of the spectrum and suggest an iterative approach to minimize the bias in the spectrum and the amplitude distribution. The null hypothesis addressed by this Iterated Amplitude Adjusted Fourier Transform (IAAFT) is that the empirical data is generated by static, invertible, nonlinear transform of a nonlinear correlated noise. In this report, the surrogates are generated by the IAAFT technique. The dynamical nonlinearities and hence the temporal structure are destroyed in the surrogate realizations. Deterministic chaos is an instance of dynamical nonlinearity and does not represent the entire class of nonlinear dynamical processes. Other instances of dynamical nonlinearities include data obtained by dynamical nonlinear transform of linearly correlated noise. Thus rejecting the null does not imply that a nonlinear deterministic dynamical process generates the given data. Some of the pitfalls and critical analysis of the surrogate algorithms can be found in [Schreiber and Schmitz, 2000, Rapp et al, 2001]. The susceptibility of the surrogate techniques to nonstationary modulated processes [Timmer, 1998] and possible methods to overcome them, based on certain assumptions is discussed elsewhere [Schmitz and Schreiber, 2000].

## 3.2 Tests for Statistical Significance

### 3.2.1 Estimating significance using traditional approach

*Parametric Testing*

Theiler et al, 1992 suggested a parametric test, which uses *sigmas* (S) for determining the statistical significance of the surrogate testing. The sigmas is defined as $S = \dfrac{\mid D_o - \boldsymbol{m}_s \mid}{\boldsymbol{s}_s}$, where $D_o$ is the value of the discriminant statistic obtained on the empirical data, $\boldsymbol{m}_s$ and $\boldsymbol{s}_s$ are the mean and the standard deviation of the discriminant statistic estimated on the ($N_{surr}$) surrogate realizations; $S$ is expressed as a factor of $\boldsymbol{s}_s$ and hence the term sigmas. The definition of sigmas can be identified with that of the z-score. The z-score expresses the deviation from the sample mean, in terms of the standard deviation. For a normally distributed data 95% of the objects have a z-score lesser than two and therefore not significant, ($\alpha = 0.05$). Thus a value of $S \geq 2$ implies that the null hypothesis can be rejected at 95% confidence. The parametric approach implicitly assumes that the distribution of the discriminant statistic obtained on the surrogate realizations is normally distributed, which might not be true, hence a drawback.

*Non-parametric testing*

A non-parametric approach was suggested [Hope, 1968; Theiler & Prichard, 1995; Rapp et al, 1994; Schreiber and Schmitz, 2000], where the number of surrogate realizations is generated based on the significance level ($\alpha$). For one-sided test, $N_{surr} = 1/\alpha - 1$ surrogate realizations are generated. The null hypothesis is rejected at $\alpha$, if the discriminant statistic on the empirical data is lesser than those on the $N_{surr} = 1/\alpha - 1$, surrogate realizations. In this report, 99 surrogates were generated corresponding to $\alpha = 0.01$.

*3.2.2 Direct statistical comparison of the distribution of the local dispersion obtained on the empirical data ( $f_e$ ) to those obtained on a single surrogate realization ( $f_s$ )*

The traditional method to determine statistical significance (Sec. 3.2.1) compares the discriminant statistic obtained on the original data to those obtained on an ensemble of surrogate realizations. In this section, we propose a direct comparison of the distribution of the local dispersions (Sec. 4) obtained on the empirical data ( $f_e$ ) to those distribution

on a *single* surrogate realization $f_s$ . As noted earlier, the empirical data is assumed to be

representative of the population. Dissipative dynamical systems settle down on an

attractor as the system evolves and are characterized by their invariant probability

measures. Therefore, an empirical sample obtained from such a system after discarding

initial transient behavior is representative of the systems dynamics and hence the

population. The surrogate realizations were generated by the IAAFT technique hence

retained the amplitude distribution and the power spectrum of the original data. This

raises the issue whether a paired statistical test is appropriate, since the data sets are not

strictly independent. However, the local dispersions, Sec. 4.2, obtained on the surrogate

cannot be paired to those obtained on the empirical data. Therefore, the choice of

unpaired test is justified. The statistical inference would be based on three separate

statistical tests: *one-sided parametric test* (ttest), *non-parametric test* (Wilcoxon-

ranksum) and *resampling with randomization*. The ttest implicitly assumes

that $f_e$ and $f_s$ are normally distributed and hence biased. The Wilcoxon-ranksum is a non-

parametric test, however it determines the statistical significance based on the ranks of

the observations and not the actual values. This can possibly introduce bias. The

resampling with randomization does not rely on any existing theoretical distributions for

determining statistical significance. However it implicitly assumes the given sample to be

representative of the population. The null hypothesis addressed in the case of the

Wilcoxon-ranksum test is that there is no significant difference between $f_e$ and $f_s$ . The alternative hypothesis addressed by the one-sided ttest and resampling with randomization is that the mean local dispersion obtained on the IAAFT surrogate is greater than that obtained on the empirical sample.

The p-value in the case of resampling with randomization was determined as follows. The difference in the mean ($\boldsymbol{m}_o$), Sec. 4, was computed on the original distributions $f_e$ and $f_s$, and on the random shuffled versions $f_e^*$ and $f_s^*$ ($\boldsymbol{m}_s^i$). For two groups containing 20 elements each, the number of possible combinations is $^{40}C_{20}$ $= \dfrac{40!}{20!20!} \sim 10^{11}$ in the case of an exact test. However, in the present study, only $i = 1...999$ random shuffles were generated for ($n_c = 20$) hence biased. The p-value obtained for the one-side test, is given by:

$$p = \frac{\text{number of cases where } (\boldsymbol{m}_s^i \geq \boldsymbol{m}_o)}{1000} \qquad (4)$$

The equal sign in the numerator of expression (4) corresponds to the original pair $f_e$ and $f_s$. A p-value lesser than 0.05, implies that the null can be rejected at significance level, $\alpha = 0.05$. The one-sided test is justified as we expect the total dispersion (Sec. 4) or the mean dispersion of the surrogate realization to be greater than that of the empirical data. For the case where each group contained eight elements (Sec. 4.1) an exact test was used to determine the p-value. The following section discusses phase-space partitioning and local dispersion estimation.

## 4. Estimating local dispersion by phase space partitioning

Partitioning of the points in the phase space, distributes the points to non-overlapping sets, $C_i, i = 1..n_c$ such that:

    *i. Non-empty*: There are no empty sets, i.e. $C_i \neq \boldsymbol{f}$

    *ii. Exclusive*: Each point is a member of only one set, i.e. $C_i \bigcap C_j = \boldsymbol{f}$ for $i \neq j$

    *iii. Exhaustive*: Each point in the phase space is a member of a set, i.e. $\bigcup_{i=1}^{n_c} C_i = U$

Constraints (i, ii, and iii) together defines a *partition*.

The k-means technique had been used in the past to partition the phase space into a finite number of clusters [Nagarajan, 2003]. The distance between any two points $(x, y)$ in a $d$-dimensional space is given by the Minkowski's metric, $\boldsymbol{d}_d(x, y) = [\sum_{i=1}^{d} |x_i - y_i|^b]^{1/b}$. The $l_2$-norm ($b = 2$, Euclidean distance) and the $l_1$-norm ($b = 1$, Manhattan distance) are special cases of the Minkowski's metric used extensively in clustering techniques. In this report, the $l_2$-norm is used as the metric however, other choices are also possible. The k-means technique generates an optimal partition by minimizing the mean-squared error [Tou and Gonzalez, 1974]. The one-dimensional data $\{w_n\}, n = 1...N$, is normalized to zero mean and unit variance. Normalization ensures the clustering to be invariant to scaling and offset errors without altering the topological structure of the attractor. The phase space reconstruction $v_q \ 1 \leq q \leq N - (d-1)\boldsymbol{t}$ of the given data is generated with embedding dimension $d$ and time delay $\boldsymbol{t}$ (1). The k-means clustering technique begins by choosing $n_c$ initial centroids. There have been several reports which address the choice of the initial centroids [Fayyad et al, 1998; Meila and Heckerman, 1998]. The ad-hoc approach would be random initialization. The splitting approach, is an iterative procedure involves generating new centroids by perturbing the given initial centroid by a factor $\in$, to obtain two new centroids [Linde et al, 1980]. Locating the modes of the distribution was also suggested for the same. [Fayyad et al, 1998]. In this report, 10% of the samples were chosen randomly and clustered. The centroids obtained were subsequently assigned as the initial guess for clustering the entire data.

## 4.1 Partitioning using k-means technique

*Step1* Choose $n_c$ initial cluster centroids as described above, represented by

$$z_1(1), z_2(1), \ldots, z_{n_c}(1).$$

*Step2* Distribute the points $v_q$ in the phase space to clusters $C_j$, $j = 1..n_c$ at the

$k^{th}$ iteration as $v_q \in C_j(k)$ if $|v_q - z_j(k)| < |v_q - z_i(k)|$, $i = 1...n_c$, $i \neq j$

*Step3* Compute the new centroids $z_j(k+1)$, $j = 1...n_c$ of the j$^{th}$ cluster by minimizing the objective function $D = \sum_{l=1}^{n_c} \sum_{z \in C_l(k)} |v_q - z|^2$.

*Step 4* Condition for the convergence of the cluster centroids.

$$\text{Is} \quad |z_j(k+1) - z_j(k)| < 0.0001 \quad, \quad j = 1,2,\ldots,n_c$$

If true stop else go to step 2.

The phase space partitioning is demonstrated by the following examples. In the subsequent sections, the eigen-values in each of the clusters are ranked, such that $l_1 > l_2 > \ldots > l_d$.

*Logistic Map*:

Consider the chaotic logistic map (F1) with ($r = 4$, $d$=2, $t = 1$, N= 2048), the SVD of the entire phase space, resulted in normalized eigen-values $l_1^* \sim 0.51$ and $l_2^* \sim 0.49$, Fig. 6a. While the logistic map is essentially one-dimensional, the eigen-values have comparable magnitudes ($l_1^* \sim l_2^*$) along orthogonal directions, suggestive of a possible two-dimensional structure, hence misleading. The equal variances can be attributed to the nonlinear correlations (parabolic shape) of the phase-space representation, Fig. 6a. A similar behavior was observed in the case of the IAAFT surrogate $l_1^* \sim 0.51$ and $l_2^* \sim 0.49$, Fig. 6b. Unlike the case of the logistic map, the equal variance observed in the surrogate can be attributed to the fact that the points fill the entire phase space, Fig. 6b hence justified. Thus, global SVD of the phase space fails to discriminate the logistic map from its IAAFT surrogate. It is important to remember that the power spectrum of the original data is retained in the surrogate realizations, which might also explain the comparable magnitude of the eigen-values obtained on global SVD (subspace decomposition) of the trajectory matrix, Section 2.1.

An eight-cluster partition ($n_c = 8$) was generated using the k-means technique for the logistic map and its IAAFT surrogate, shown in Fig. 6a and Fig. 6b respectively. While $l_1^*$ represents the component along the curve, $l_2^*$ represents that component perpendicular to it. The non-zero value of $l_2^*$ is solely due to the nonlinear effect. The relative magnitude of $l_1^*$ increased with respect to $l_2^*$ with increasing number of clusters. Thus increasing the number of clusters minimizes the nonlinear effects. For instance, in cluster $C_6$, $I_1^{6*} \sim 0.8$ and $I_2^{6*} \sim 0.2$, Fig. 6a, similar distribution in the magnitude of the eigen-values were observed for the other clusters $C_i$, $i = 1...8$. The distribution of the eigen-values ($I_1^i, I_2^i$) for the logistic map and its IAAFT surrogate is shown in Fig. 6c.

*Henon Map*:

For the chaotic henon map (F2) with ($a = 1.4$, $b = 0.3$, $d = 2$, $t = 1$, N= 2048), the SVD of the entire phase space, resulted in $I_1^* \sim 0.57$ and $I_2^* \sim 0.43$, Fig. 7a, comparable to that obtained on the surrogate realization $I_1^* \sim 0.57$ and $I_2^* \sim 0.43$, Fig. 7b. Unlike the case of the logisitic map, the manifold of the henon map is quite complex, Fig. 7a. Therefore, an eight-cluster partition may not be sufficient to minimize the nonlinear effects. This is evident from the distribution of the points among the clusters, Fig. 7a. The choice of the number of clusters can be crucial in the statistical discrimination and is discussed in the following section. The phase space partition of an IAAFT surrogate of the henon map is shown in Fig. 7b. The distribution of the eigen-values for an eight-cluster partition of the henon map and its IAAFT surrogate realization is shown in Fig. 7c.

## 4.2 Local Dispersion

The local and dispersions of a given data is determined as follows:

Step 1: Normalize the data to zero mean unit variance.

Step 2: Embed the data with embedding dimension ($d$) and time delay ($t$).

Step 3: Partition the embedded data into $n_c$ clusters.

Step 4: Let the $N_i$ vectors in the $i^{th}$ cluster of a $d$-dimensional space be represented by the local trajectory matrix $\Gamma^i_{N_i \times d}$. Assuming that $\Gamma^i$ is locally linear, the SVD of $\Gamma^i$ can be used to extract the $d$ singular values.

Step 5: The local dispersion associated with the $i^{th}$ cluster is given by the product of the eigen-values along the $d$ dimensions, i.e.

$$\Delta^i = \prod_{j=1}^{d} \boldsymbol{l}^i_j \tag{4}$$

As noted earlier, Sec. 2, the eigen-values is proportional to the axes of a $d$-dimensional ellipsoid. In which case, $\Delta^i$ would represent the volume of the $i^{th}$ cluster.

Step 6: The total dispersion $\Delta^T$ of the entire phase space is given by the sum of the local dispersions, i.e.

$$\Delta^T = \sum_{i=1}^{n_c} \Delta^i = \sum_{i=1}^{n_c} \det(\Gamma^{i^T} \Gamma^i)^{1/2} \neq \det(\sum_{i=1}^{n_c} (\Gamma^{i^T} \Gamma^i)^{1/2}) \tag{5}$$

Relation (5) is direct application of the properties of determinants. Therefore, $\Delta^T$ is a *nonlinear statistic*.

### 4.2.1 Importance of eigen-values comprising the noise floor in statistical discrimination

Determining the dominant eigen-values by SVD of $\Gamma^i$ can be used to estimate the local dimensionality. One possible approach would be to choose those eigen-values greater than or equal to a predefined threshold $\boldsymbol{q}$, i.e. $\boldsymbol{l}^i_j \geq \boldsymbol{q}$, $j = 1..d$. While the eigen-values $\boldsymbol{l}^i_j$, $j = 1..d$ are uniformly distributed in the case of white noise, they are likely to be skewed with increasing correlation. The log transform is widely used in statistical literature to minimize the skew in the distributions. For the $i^{th}$ cluster, the mean of the log-transformed eigen-values is given by

$$\log(\boldsymbol{q}) = \frac{1}{d} \sum_{j=1}^{d} \log(\boldsymbol{l}_j^i) = \log(\prod_{j=1}^{d} \boldsymbol{l}_j^i)^{1/d}$$

The above expression reduces to the geometric mean

$$\boldsymbol{q} = (\prod_{j=1}^{d} \boldsymbol{l}_j^i)^{1/d} \tag{6}$$

The geometric mean is less susceptible to outliers compared to the arithmetic mean, and also represents the average generalized variance of a multivariate process [Jobson, 1996]. Thus a *crude* estimate of the dimension of $\Gamma^i$ would be equal to the number of eigen-values satisfying the constraint $\boldsymbol{l}_j^i \geq \boldsymbol{q}$, $j = 1..d$, with $\boldsymbol{l}_j^i < \boldsymbol{q}$ representing the *noise floor*. The first report, to our knowledge that addressed the issue of intrinsic dimensionality was by Fukunaga and Olsen, 1971. They demonstrated the pitfalls of linear transformation techniques when applied to nonlinear functions of random variables, and suggested a local approach to estimate the intrinsic dimensionality. [Passamante et al, 1989] suggested fixed neighborhood size criteria, so that the assumption of locally linear subspace is valid. While the dominant eigen-values may be useful in determining the local intrinsic dimension, the eigen-values corresponding to the noise floor play an important role in the statistical discrimination between the given data and its surrogate realization, and shall be demonstrated by the following examples.

*a. Nonlinear Deterministic Data*

The distributions obtained on the given empirical data and the surrogates were tested for statistical significance using three different tests, namely: parametric, non-parametric and permutation test. Parametric one-side ttest performs the test against the alternative hypothesis that the mean of the local dispersions obtained on the empirical data is lesser than the mean of the local dispersions obtained on the surrogates. Since the number of clusters is fixed at ($n_c$ = 7), all possible permutations was generated for the one-sided resampling with randomization (i.e. $^{14}C_7$ = 3432). The two-sided ranksum was used to address the null that the median local dispersions of the original data is not significantly different from the median local dispersions on the surrogate.

*Logistic Map*

A ($n_c$ = 8) partition of the logistic map (F1) ($d$ =2, $t$ = 1, N = 2048, r = 4.0) and its IAAFT surrogate was generated. The SVD of each of the clusters in the two-dimensional space resulted in two eigen-values, ($l^M = l^i_1$) as dominant and ($l^L = l^i_2$) representing the noise floor, i.e. $l^M > q \geq l^L$, where $q$ is the geometric mean (7). The distribution of the eigen-values for the logistic map and its IAAFT surrogate is shown in Fig. 6c. While the distribution of the dominant eigen-value showed a significant overlap between the empirical data and its surrogate, the distribution of were non-overlapping. As noted in Sec. 4.1, increasing the number of clusters increases the relative magnitude of tangential component $l^i_1$ with respect to the normal component $l^i_2$. However, the value of $l^i_1$ obtained on the surrogate is on average greater than that of the logistic map. This can be attributed to the fact the geometry of the phase space is destroyed in the surrogates as opposed to that of the empirical data. Direct statistical comparison of the distribution of $l^i_1$ corresponding to the dominant eigen-value, between the empirical data, $f_e$ and its IAAFT surrogate, $f_s$ failed to reject the null, Table I. This was confirmed by parametric, non-parametric and randomization tests. Interestingly enough, statistical testing of the distribution of $l_2^i$ corresponding to the noise floor rejected the null $(p < 10^{-2})$, Table I. This can be attributed to the non-overlapping distributions of $l_2^i$ obtained on the empirical data and its surrogate realization, shown in Fig. 6c.

*Henon Map*

A similar behavior was observed for a $(n_c = 8)$ partition of the chaotic henon map (F2) with $(d = 2, t = 1, N = 2048, a = 1.4, b = 0.3)$. The distribution of eigen-values obtained on the henon data and its IAAFT surrogate realization are shown in Fig. 7c. While the distribution of the dominant eigen-values $(l^M = l_1^i)$ was not statistically significantly, those corresponding to the noise floor $(l^L = l_2^i)$ were significant $(p < 10^{-2})$ Fig. 7c. The corresponding p-values, obtained using the parametric, non-parametric and resampling statistic is enclosed in Table I A similar behavior was observed in the case of the bleached henon map. The FNN (Sec. 1) failed to converge in the case of the bleached henon data, hence the embedding parameters were chosen as that of the original henon data, i.e. $(d = 2, t = 1)$, Table I.

*Chaotic Laser*

The laser data [Huebner et al, 1989] was embedded with parameters $(d = 7, t = 1)$. The FNN algorithm (Sec. 1) was used to determine the optimal embedding dimension. Statistical testing failed to reject the null for $(l^M = l_1^i)$ but it rejected the null for $(l^L = l_7^i)$ $(p < 10^{-2})$, Table I . Recent literature has pointed out the susceptibility of discriminant measure to the presence of noise [Schreiber & Schmitz, 1997]. Noise can be broadly classified into observational (extrinsic) of the form $y_t = f(x_t) + \in_t$ and dynamical (intrinsic) noise $x_{t+1} = f(x_t + \in_t)$. The latter is coupled to the dynamics and its effects are non-trivial. Noise can further be classified into in-band and out-band noise [Schreiber & Schmitz, 1997]. Unlike out-band noise, in-band noise (Appendix, F10) retains the power spectrum of the given data. This in turn would imply that subspace decomposition of the in-band noise and the given data would yield the same eigen-spectrum (Sec. 2), their magnitude being proportional to the noise factor (*a*), see F10. However, this is not true in the case of out-band noise. A recent study [Schreiber & Schmitz, 1997], addressed the effect of in-band noise on the statistical power of the various discriminant measures with increasing noise factor *a*, (F10). We generated laser data with inband noise using the same parameters as in [Schreiber and Schmitz, 1997, F10]. The tests failed to reject the null for $(\boldsymbol{l}^M = \boldsymbol{l}_1^{i*})$ and $(\boldsymbol{l}^L = \boldsymbol{l}_7^{i*})$. This can be attributed to the distortion introduced in the phase space which is sensitive to the number of clusters and results in an increased overlap between the empirical data, $f_e$ and the IAAFT surrogate $f_s$.

### b. Non-deterministic Data

A ($n_c = 8$) partition was generated for the nonlinearly correlated noise (F7) with ($d = 2$, $\boldsymbol{t}$ = 1, N= 2048). The distribution of eigen-values $(\boldsymbol{l}^M = \boldsymbol{l}_1^i)$ and $(\boldsymbol{l}^L = \boldsymbol{l}_2^i)$ obtained on the given empirical data and its IAAFT surrogate realization are shown in Fig. 8a and 8b respectively. Unlike the case of the chaotic data sets, the phase space empirical data does not exhibit any characteristic structure. This is reflected in similar distribution of the points in the phase space in the empirical data and its IAAFT surrogate. The distribution of the dominant eigen-value $(\lambda^M = \lambda^i_1)$ and those corresponding to the noise-floor $(\lambda^L = \lambda^i_2)$ exhibited a significant overlap, Fig. 8c. Statistical testing revealed no significant difference in either case, see Table I. As opposed to dynamical nonlinearities in the above cases, static nonlinearities are retained in the surrogates by rank-ordering. Similar analysis in the case of linearly correlated noise and white noise ($d = 2$, $\tau = 1$, N = 2048) failed to reject the null hypothesis across the three statistical tests for $(\lambda^M = \lambda^i_1)$ and $(\lambda^L = \lambda^i_2)$, Table I.

From the above case studies, it is clear that the eigen-values comprising the noise floor cannot be disregarded in statistical discrimination. This in turn implies that the choice of the threshold $q$ can have a prominent effect the results. One possible solution would be to retain the entire spectrum of eigen-values. However, the results obtained by retaining the entire spectrum are dependent on the choice of the number of clusters. This is especially true in the case of nonlinear deterministic data sets. As noted earlier, the manifolds of some of the data sets are quite complex and may require more number of clusters for statistical discrimination. The number of clusters also determines the number of points in the distribution of the local dispersions and can hence affect statistical discrimination. In the above case studies, only one sample was considered to arrive at the p-values enclosed in Table I. A more practical approach would be to estimate the power for each case by generating multiple realizations. However, we believe in the absence of noise, the above results would have a power ($> 0.8$).

## 4.3 Volume Collapse - effect of increasing number of clusters

The local dispersion $\Delta^i$, of the $i^{th}$ cluster in a $d$-dimensional space is proportional to the volume of $i^{th}$ $d$-dimensional ellipsoid. Increasing the number of clusters generates finer partitions of the empirical data, and results in local subspaces with rank lower than $d$, since the annihilation of an eigen-value decreases the rank of the corresponding correlation matrix by one. For chaotic data obtained from dissipative dynamical systems, embedded in a finite-dimensional phase space we expect the eigen-value along the $j^{th}$ axis of the $i^{th}$ ellipsoid tends to zero, i.e. $l_j^i \rightarrow 0$, with increasing $n_c$, as a result $\Delta^i \rightarrow 0$, this shall be referred to as *volume collapse*. The surrogate realizations are constrained randomized versions of the original data and hence cannot be embedded in a finite dimensional space. Thus the volume collapse observed on the surrogate data is *on an average*, *lesser* than that of the empirical data. Alternatively, the total dispersion $(\Delta^T)$ of the surrogate realization is *on an average greater* than that of the empirical data. In the subsequent section, the total dispersion along with the IAAFT surrogate shall be used as a discriminant statistic in determining the nature of the nonlinear correlation in the given data.

### 4.3.1 Choice of the number of clusters

The k-means partitioning technique divides a given data into a specified number of clusters. There have been reports that address the issue of cluster size based on class separability [Tou and Gonzalez, 1974]. It should be noted that the objective here is to choose the number of clusters, so as to minimize the nonlinear effects and not to separate the given data into distinct classes. As noted earlier in, Sec. 2.2, nonlinearities can arise in deterministic as well as non-deterministic settings. The IAAFT surrogates, generated by constrained randomized shuffle of the given empirical data were used as reference data sets to determine the number of clusters. We use the index [Nagarajan, 2004], given by

$$r(n_c) = 1 - \Delta^{T_0}(n_c) / \Delta^{T_s}(n_c) \qquad (7)$$

where $\Delta^{T_0}$ and $\Delta^{T_s}$, represent the value of the total dispersion, obtained on the empirical data and its IAAFT surrogate to determine $n_c$. If the empirical data were generated by the null hypothesis (i.e. static, invertible nonlinear transform of a linearly correlated noise), one would fail to observe a discrepancy between $\Delta^{T_0}$ and $\Delta^{T_s}$ with increasing $n_c$ and the value of $r(n_c)$ would fluctuate around zero, i.e. $\Delta^{T_0}(n_c) \sim \Delta^{T_s}(n_c)$ or $r(n_c) \sim 0$. This is due to the fact the static invertible nonlinearites are retained in the surrogates by rank-ordering and the linear correlations are preserved through phase randomization. Thus any choice of $(n_c)$ would fail to show statistical significance, hence a valid choice. However, for chaotic data sets (F1, F2, F8, F9, F10), a proper choice of $n_c$ is necessary for statistical discrimination. Spurious results can arise in extreme cases: when the number of clusters is comparable to the number of points in the phase space, i.e. $n_c \sim$ N-$(d$-$1)t$ and for the singleton cluster, i.e. $n_c = 1$. In the former, failure to reject the null is due to inadequate number of points (density) to capture the dynamics in a given partition, while the latter represents global SVD and is bound to fail in the presence of nonlinear correlations, Sec. 2. The required number of clusters for statistical discrimination lies between these extremes. For chaotic data sets, one observes volume collapse with increasing $(n_c)$, resulting in increased discrepancy between $\Delta^{T_0}$ and $\Delta^{T_s}$. The variation of $r(n_c)$ with $n_c$, for data sets generated by chaotic and non-deterministic processes are shown in Fig. 9 and Fig. 10 respectively. The chaotic data sets considered include (F2, F8, F9 and F10).

In Sec. 2.1, we demonstrated how linear filtering of chaotic data could introduce characteristic distortions in the phase space. The susceptibility of certain discriminant statistics to linear filtering also termed as bleaching was reported earlier by [Theiler and Eubank, 1993]. In the case of experimental data, linear filtering is an integral part of the data acquisition and may be unavoidable. The bleached henon data (F9) was generated as described in [Theiler and Eubank, 1993].

While the plot of $r(n_c)$ versus $n_c$ increases in the case of chaotic processes with a plateau around ($n_c \sim 20$), it fluctuates around zero in the case of non-deterministic processes, Figs. 9 and 10. Thus for the subsequent analysis we fixed the number of clusters, $n_c = 20$. The variance about the mean for 100 individual realizations is shown in Figs. 9 and 10. The individual realizations were generated by clustering the data with different initial conditions. This was done in order to accommodate the sensitivity of the clustering technique. The approximate entropy (ApEn) had been shown to be an effective discriminant statistic [Pincus, 1991]. The ApEn was computed on the original data (ApEn$^o$) and its corresponding surrogate realization (ApEn$^s$). The ratio $r$(ApEn) = 1 − ApEn$^o$/ApEn$^s$, obtained for the nonlinear deterministic and non-determinisitic data sets is shown for reference in Figs. 9 and 10 respectively. The spurious deviation from zero, observed in the case of ARMA process can be attributed to the smearing of the spectral power across the frequency band in the surrogate realizations, Sec. 2, Fig. 4.2. Both the measures are susceptible to ARMA processes. Thus it is possible to identify nonlinear structure in the case of narrow band ARMA process, also reflected by the high value of $r(n_c)$ and $r$(ApEn) . The effect of noise dominates the dynamics in the case of the laser with inband noise (F10), Fig. 9d, this is reflected by the low value of $r(n_c)$ and $r$(ApEn).

Also shown for reference is $r$ estimates obtained using Morgera's covariance complexity (Appendix, E) as a discriminant measure. The covariance complexity $h$ of the entire phase space is estimated in a manner similar to the total dispersion $D^T$, Sec. 4.2, (Appendix E). Similar to $D^T$ the discrepancy in $h$ between the empirical data and its IAAFT surrogate increases with increasing number of clusters for the henon map Fig. 11, whereas no prominent increase is observed in the case of the linearly correlated noise, Fig. 11. This is also reflected in the index $r$, Fig. 11.

*a. Direct statistical comparison of the distribution of the local dispersions*

The distribution of the local dispersions obtained on the empirical data was compared directly to those obtained on an IAAFT surrogate using three different statistical tests, Sec. 3. The cluster size was chosen as $n_c = 20$, Fig. 9 and Fig. 10. The null hypothesis was rejected in the case of chaotic logistic, henon and laser data. The local dispersions obtained on the and bleached henon data and laser with inband noise showed significant overlap with their IAAFT surrogates, hence the tests failed to reject the null. As expected, the tests failed to reject the null for the non-deterministic data, namely: random shuffle, linearly correlated noise and nonlinearly correlated noise.

*b. Traditional statistical discrimination using parametric and non-parametric tests*

In traditional surrogate testing, the discriminant obtained on the empirical data is compared to those obtained on an ensemble of surrogate realizations, i.e. the total dispersion obtained on the empirical data is compared to the distribution of the total dispersions obtained on the ($N_{surr}$) surrogate realizations, Sec. 3.2.1. Parametric and non-parametric tests were used to estimate the statistical significance. One hundred different realizations were generated, corresponding to ($\alpha = 0.01$). The null was rejected in the case of the nonlinear deterministic data (Table III) and not in the case of the non-deterministic data. Unlike traditional statistical discrimination of the total dispersion, direct comparison of the statistical distributions of the local dispersion is sensitive to noise.

*c. Effect of the embedding dimension on* $\mathbf{r}(n_c)$

Techniques such as FNN are susceptible to the noise and the finite length of the data set and may fail to converge. Thus it might be important to investigate the effect of varying embedding dimension and number of clusters on the $\mathbf{r}$ estimates. For the linear and nonlinearly correlated noise, the embedding dimension was varied $d = 2, 4, 6, 8$ and $10$, N = 2048, $\mathbf{t} = 1$) and the number of clusters ($n_c = 2…20$). The $\mathbf{r}$ for each ($d$, $n_c$) was obtained by averaging over ten realization, Fig. 11. As expected, the value of $\mathbf{r}$ fluctuates around the value zero, with varying parameters indicating possible non-deterministic nature of the linear and nonlinearly correlated noise, shown in Fig 11a and 11b respectively. Thus varying embedding dimension and number of clusters do not affect the results in the case of non-deterministic data. However, this is not so in the case of nonlinear deterministic data. Consider the chaotic laser and henon data corrupted with inband noise (F10). The FNN algorithm failed to converge for these data sets hence the embedding dimension could not be determined. The choice of ($d = 7$, $t = 1$) for the inband laser resulted in $r$ estimates ($\sim 0.2$), Fig. 9. Such low $r$ values were also observed in the case of the chaotic henon data with inband noise. The variation of $r$ averaged over ten realizations with varying embedding dimension $d = 2, 5, 7, 10, 12$ and $14$, N = 2048, $t = 1$) and the number of clusters ($n_c = 2\ldots20$) is shown in Fig. 12. Unlike the non-deterministic data, there is an increase in the $r$ value with increasing embedding dimension and number of clusters. Thus embedding noisy data in a high dimensional space might be useful in statistical discrimination.

## 5. Discussion

In this report, the inherent assumptions and caveats in the application of linear transformation approaches such as SVD to nonlinear correlated data were reviewed. The close connection between the eigen decomposition and the power spectral estimation is revealed by invoking Caratheodory's uniqueness result. Spectral estimation of a data composed of harmonically related frequencies, is related to the eigen decomposition of the trajectory matrix with unit time delay ($t = 1$) and embedding dimension ($d > p$), where $p$ denotes the number of distinct frequency components in it. Thus SVD of the trajectory matrix of periodic waveform, which is essentially one-dimensional, reveals its periodicity as opposed to its dimensionality. Period doubling is a possible route to chaos. A possible explanation for the failure to observe a sharp decrease in magnitude of eigen-values in the nonlinear dynamical processes such as chaos can be attributed to its broad-band power spectrum. The false positives observed in surrogate testing of narrow band ARMA process can be attributed to the smearing of the spectral power across the frequency band, in the corresponding surrogate realization. The spectral smearing is reflected in the spurious non-zero eigen-values and discrepancy between the empirical data and its surrogate realization.

SVD of the phase space of nonlinearly correlated data can result is spurious eigen-values which are the outcome of the nonlinear effects as opposed to the dimensionality. Partitioning the phase space can be useful in minimizing the nonlinear effects. In this report, the k-means clustering technique was used to achieve the same. Other choices of partitioning can also be used. It is important to note that the nonlinear correlations can arise in the case of deterministic and non-deterministic settings. An example of the former is a chaotic process while that of the latter is a nonlinear transform of a correlated noise. While partitioning can minimize the nonlinear effects, it would be unhelpful in distinguishing the nature of the nonlinearity. In this report, surrogate testing along with partitioning was used to discriminate nonlinearities arising in deterministic and non-deterministic settings. The surrogates were generated using the iterated amplitude adjusted Fourier transform (IAAFT). The local dispersion is given by the product of the eigen-values obtained by SVD of the corresponding local trajectory matrix. The distribution of the local dispersion and the sum of the local dispersions (total dispersion) were used as discriminant statistics. The importance of the eigen-values comprising the noise floor in statistical discrimination was illustrated with several examples. Choosing eigen-values greater than a pre-defined threshold can provide a crude estimate of the intrinsic dimensionality, however they have a prominent effect on the statistical discrimination. The choice of the pre-defined threshold is non-trivial and can give rise to spurious results. Retaining the entire spectrum of eigen-values is suggested as a possible alternative. The index (*r*) is proposed to determine the number of clusters required for statistical discrimination. It relies on the fact that data obtained from nonlinear deterministic, dissipative dynamical systems, are finite dimensional and settle down on an attractor after initial transients, as opposed to the surrogate counterpart. With increasing number of clusters, the volume collapse observed on an average in the empirical data is greater than that of its surrogate realization. However, for data sets generated by static, invertible, nonlinear transforms of linearly correlated noise, the static, invertible nonlinear transform is retained in the surrogates by a rank ordering approach and the power spectrum, hence the auto-correlation, by phase randomization. Thus, increasing the number of clusters does not have a significant effect on the statistical discrimination. Any choice of clusters is valid for non-deterministic data generated by the null hypothesis.

Two different approaches were used to estimate the statistical significance. In the first approach, the distribution of the local dispersions obtained on the empirical data was statistically compared to that obtained on the surrogate data. It should be noted that in this case, a single realization of the empirical data is compared against a single realization of the surrogate data. Three tests, namely: one-sided ttest, Wilcoxon-ranksum and one-sided resampling statistics were used to estimate statistical significance. The distribution of the local dispersions is determined by the number of clusters and hence can affect the statistical discrimination. The second approach is the traditional surrogate analysis, which generates a number of surrogate realizations for a given empirical data. The total dispersion was estimated on the empirical data and the surrogate realizations. Parametric and non-parametric tests were used to determine the statistical significance. The traditional surrogate testing proved to be superior compared to the direct comparison of the distributions of the local dispersion, especially in the case of noisy data.

In the case of non-deterministic data and nonlinear deterministic data corrupted with noise, it might not be possible to estimate the embedding dimension. The variation of the index $r$ with increasing embedding dimension and number of clusters were determined for these data sets. It was found that embedding the noisy nonlinear deterministic data in high dimensional space may be useful in their statistical discrimination. There was no prominent change observed for non-deterministic data sets, indicating that the choice of the embedding dimension or the number of clusters does not affect their results. The local dispersion proposed in this report, is immune to the ordering of the vectors in the trajectory matrix. A possible extension of the above technique would be to incorporate the orientation of the vectors in the phase space. Such an approach would make the choice of the surrogate algorithm redundant. This is an issue that needs to be further investigated.

**Acknowledgements**

The IAAFT surrogates were generated using the TISEAN (TIme SEries ANalysis) routines developed by Hegger, R. Kantz, H. and Schreiber, T. We would like to thank Dr. Paul Rapp, Department of Pharmacology and Physiology, Drexel University College of Medicine for his comments and suggestions.

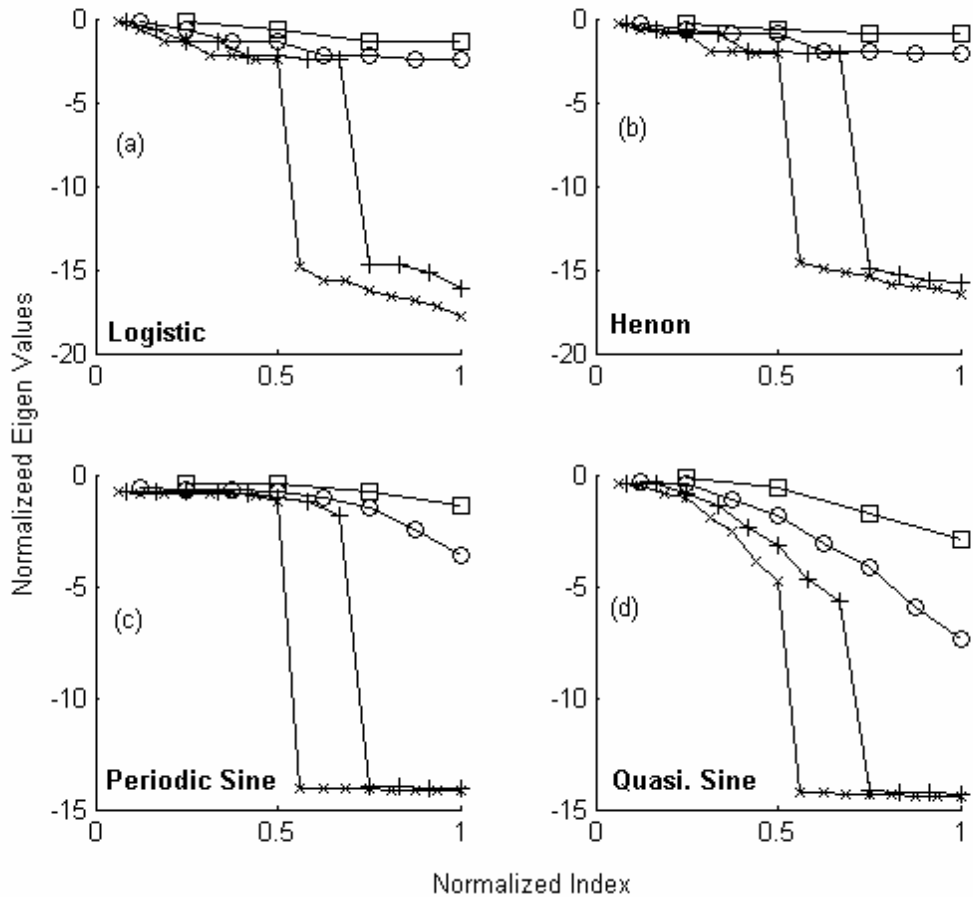

**Figure 1** Variation of the sorted normalized eigen-values $\log_{10}(\lambda^*)$ with embedding dimension $d = 4$ (square) , 8 (o), 12 (+), 16 (x), for a period eight limit cycle (N = 2048, **$t$** = 1) generated by logistic map with parameter $r = 3.55$ and henon map with parameter ($a=1.03$, $b=0.3$) is shown in (a) and (b) respectively. The variation of $\log_{10}(\lambda^*)$ for periodic (period 8) and quasi-periodic sinusoid is shown in (c) and (d) respectively. No sharp transitions were observed for $d = 4$ and 8, however, a marked decrease was observed between the eighth and ninth eigen-values for $d = 12$ and 16, reflecting the periodicity of the data in cases (a, b and c).

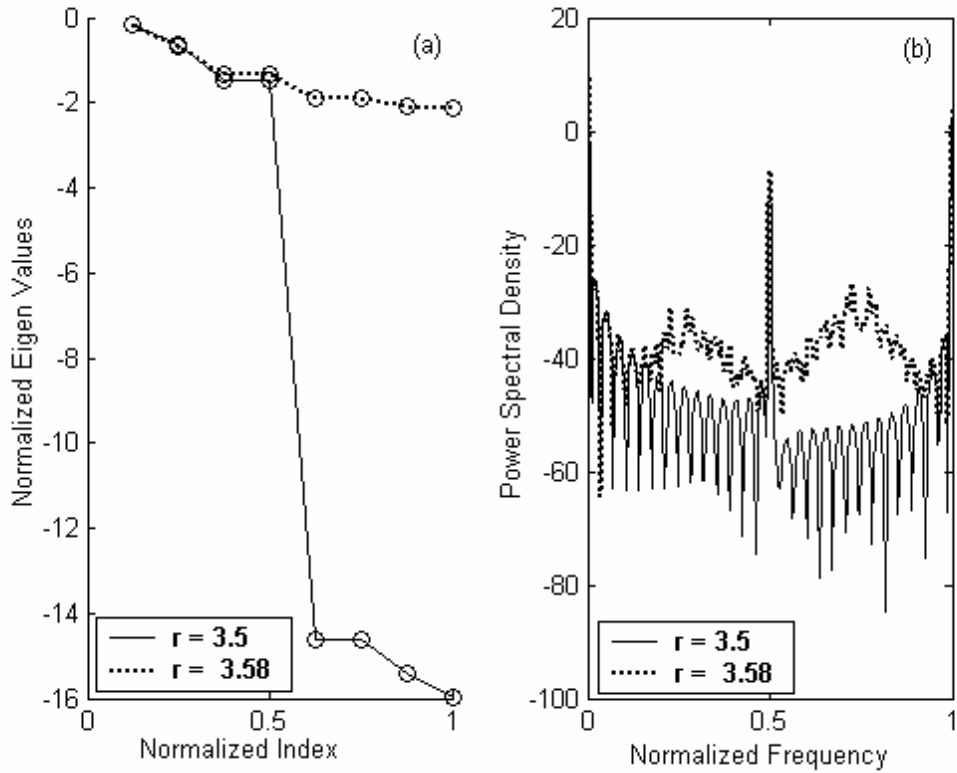

**Figure 2** Variation of the sorted normalized eigen-values $\log_{10}(\lambda^*)$ for embedding dimension $d = 8$, for data (N = 2048, $t = 1$) generated from logistic map with $r = 3.5$ (solid line) and $r = 3.58$ (dotted line). A sharp transition in $\log_{10}(\lambda^*)$ is observed at $d = 4$ for ($r = 3.5$), however no sharp transition is observed for the case ($r = 3.58$). The corresponding power spectrum is shown on the right.

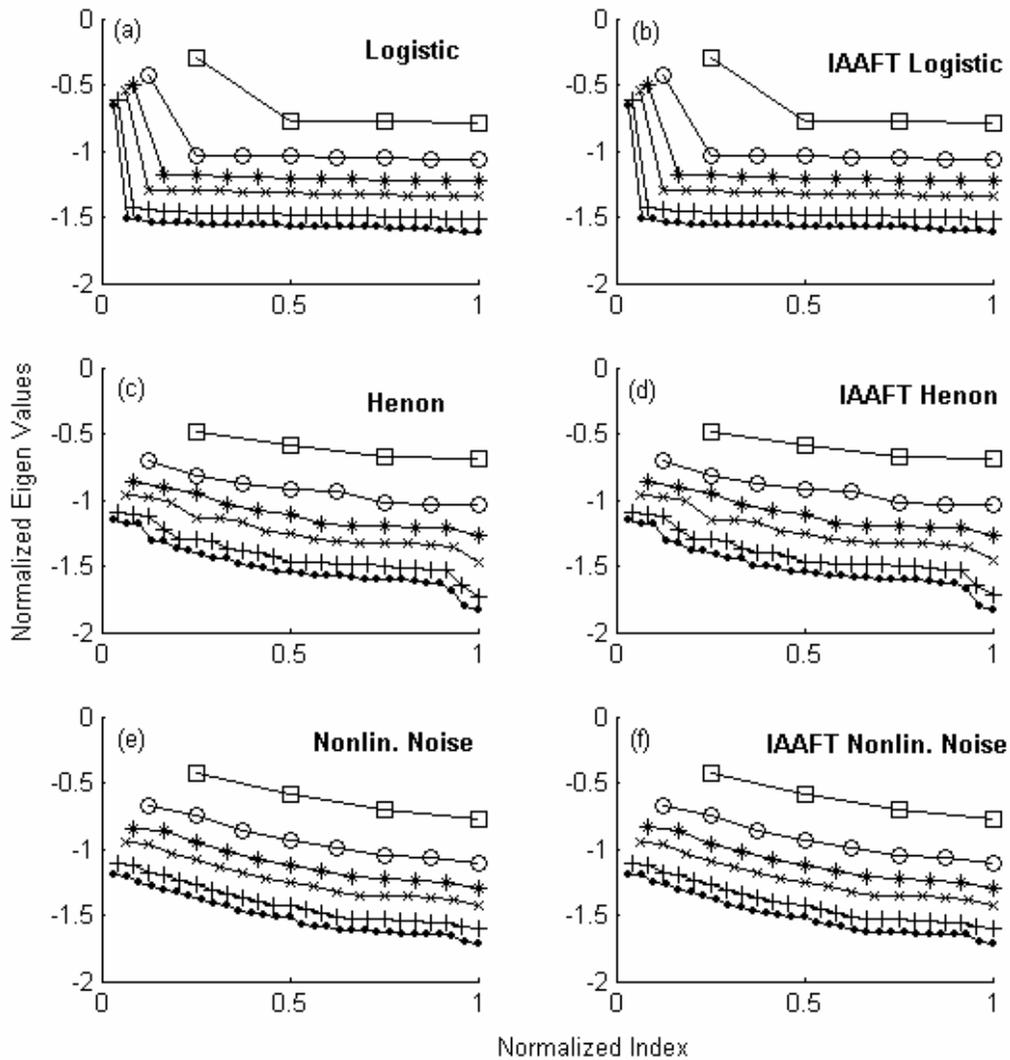

**Figure 3** Variation of the sorted normalized eigen-values $\log_{10}(\lambda^*)$ with embedding dimension $d = 4$ (square) , 8 (o), 12 (*), 16 (x), 24 (+), 30 (.) for chaotic logistic map ($r = 4.0$, N = 2048, $t = 1$), henon map (a = 1.4, b = 0.3 N = 2048, $t = 1$) and nonlinearly correlated noise (N = 2048, $t = 1$) are shown in (a), (c) and (e) respectively. The variation of the sorted normalized eigen-values on their IAAFT surrogates is shown in (b), (d) and

(f) respectively. No sharp transitions were observed for $d$ = 4, 8, 12, 16, 24, and 30 failing to reflect the dimensionality of the dynamical system.

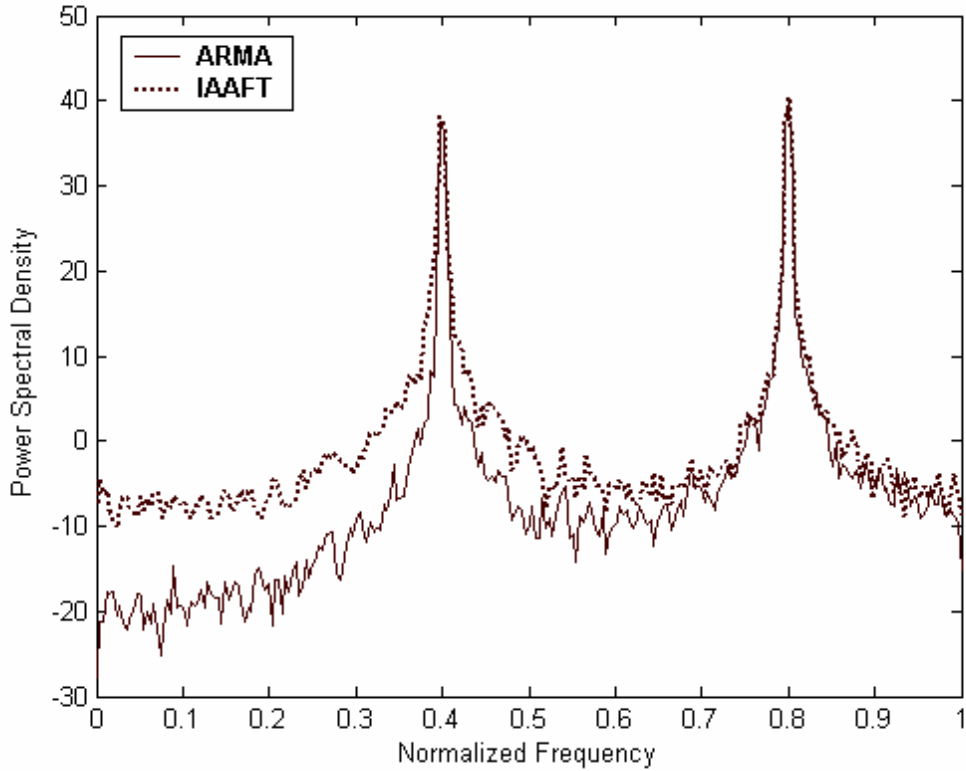

**Figure 4a** Power spectrum of the narrow band ARMA process (solid line) and that of its IAAFT surrogate (dotted line).

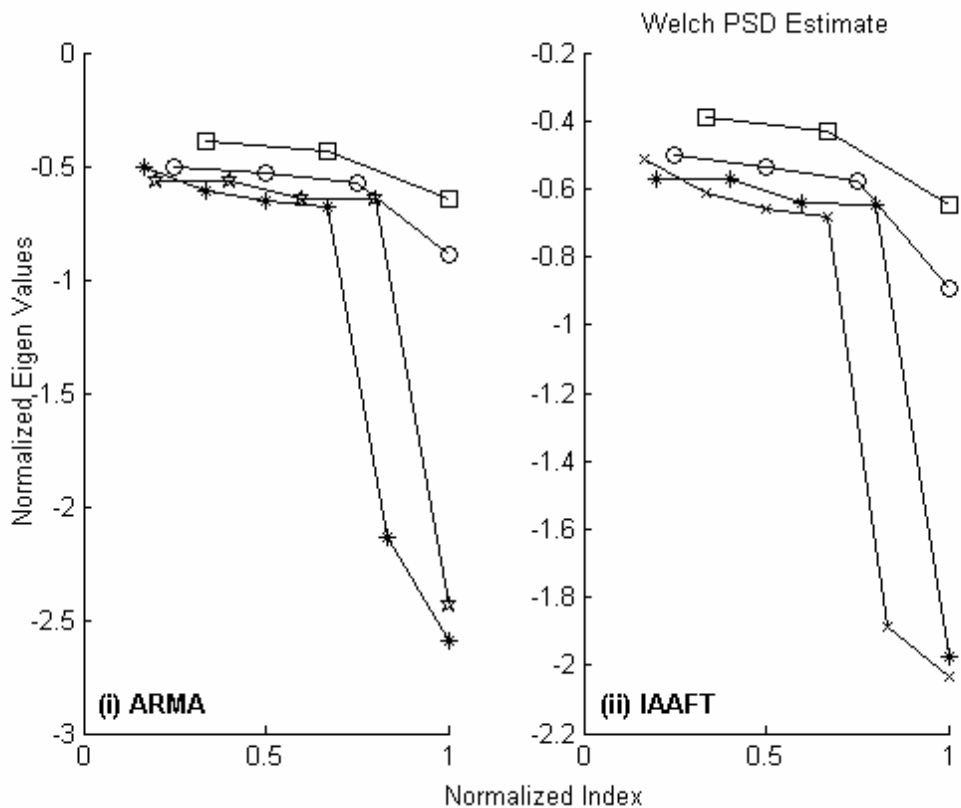

**Figure 4b** Variation of the sorted normalized eigen-values $\log_{10}(\lambda^{*})$ with embedding dimension $d = 3$ (square) ,4 (o), 5 (*)and 6 (x), (N = 2048, $\boldsymbol{t}$ = 1) for the ARMA process and that of its IAAFT surrogate is shown in (i) and (ii) respectively. Unlike (ii), a transition in the magnitude of the eigen-values was observed between $d = 4$ and $d = 5$ is observed in (i).

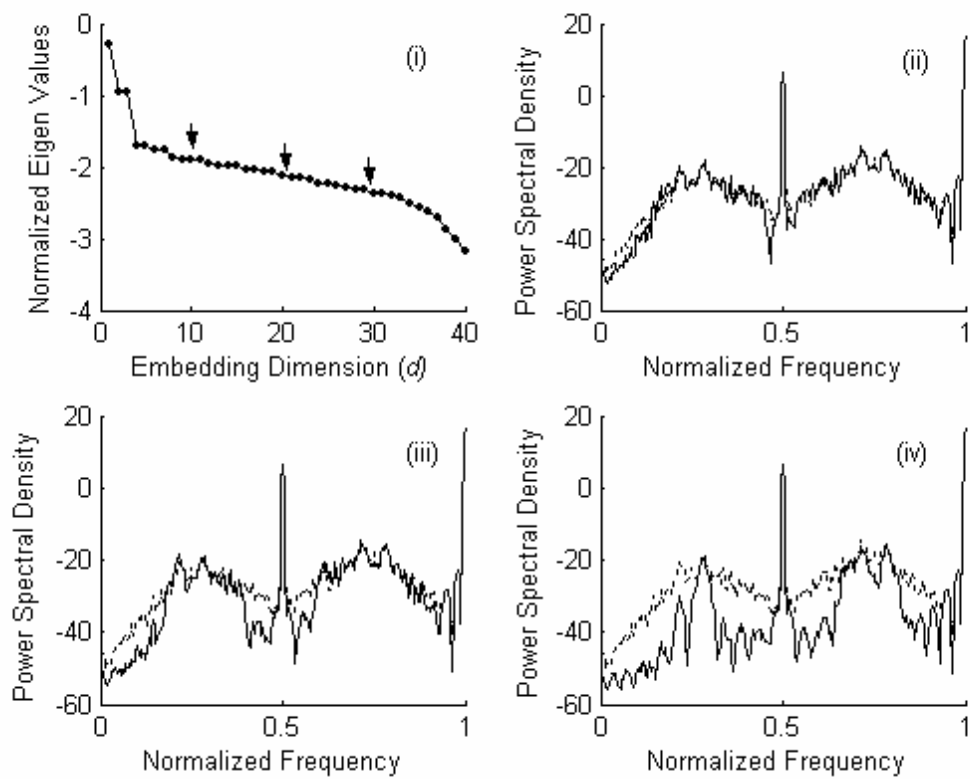

**Figure 5a** The decay in the normalized eigen-values $\log_{10}(\lambda^*)$ of the logistic map ($r = 3.58$, N = 2048) with embedding parameters (d = 40, $t$ =1) is shown in (i). The power spectrum constructed after linear filtering using the dominant (30, 20 and 10) eigen-values are shown in (ii, iii, and iv) respectively (solid lines). The power spectrum of the original data is shown for reference (dotted lines).

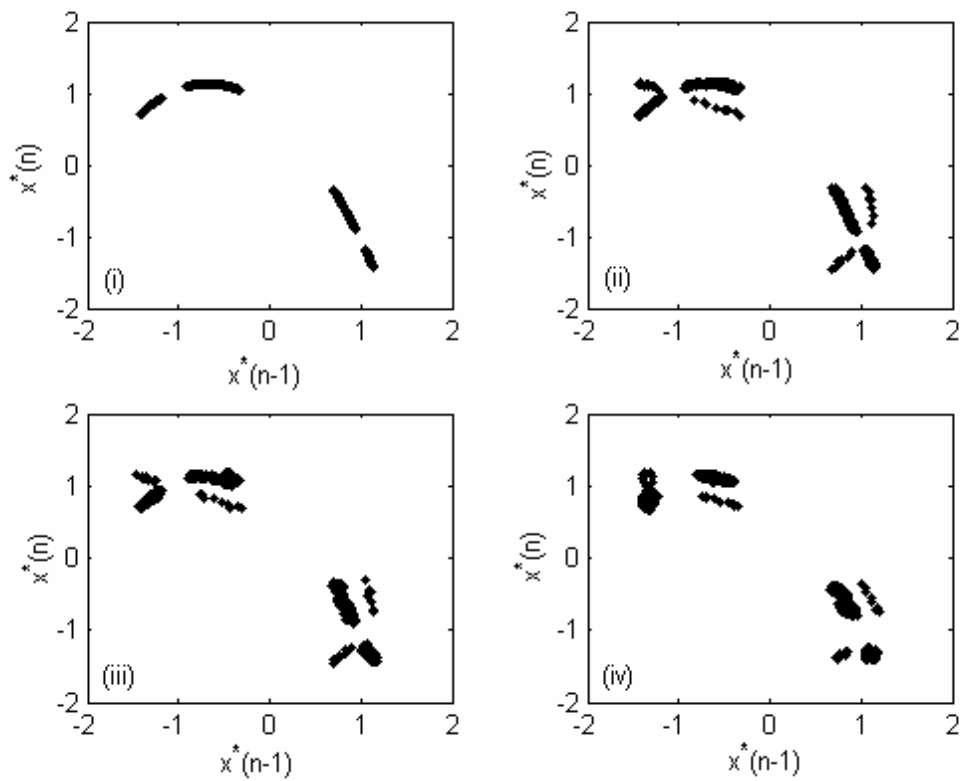

**Figure 5b** Phase space representation of the logistic map ($r$ = 3.58, N = 2048, $d$ = 2, $t$ = 1), is shown in (i), and that of its linear filtered counterparts, obtained using the dominant (30, 20 and 10) eigen-values is shown in (ii, iii and iv) respectively.

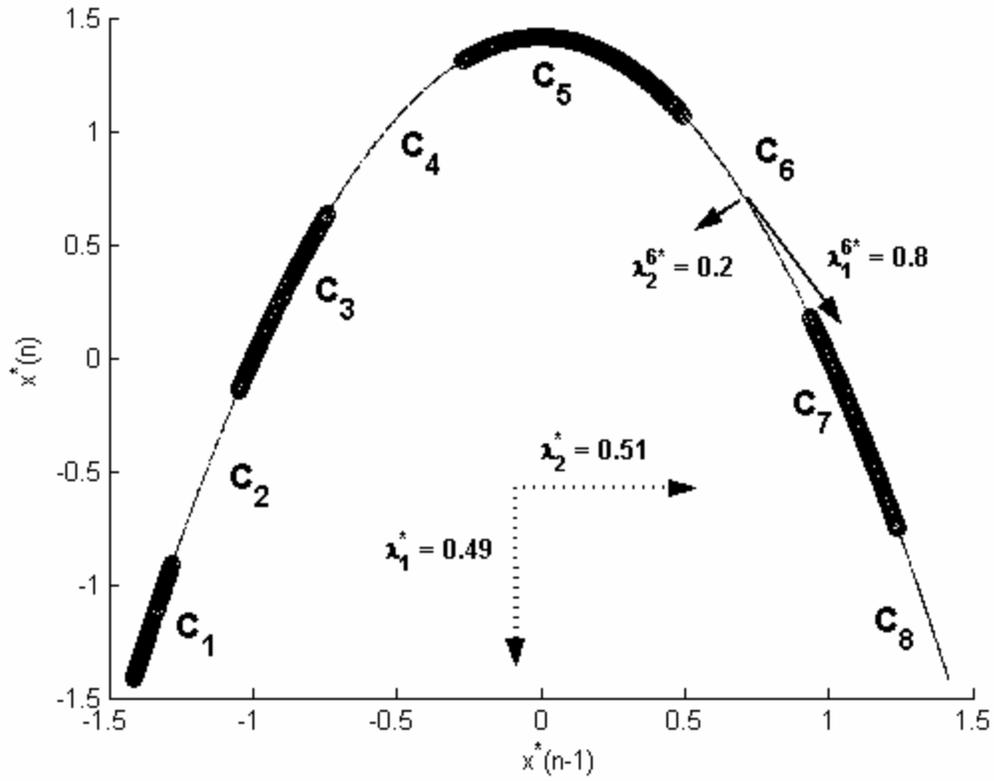

**Figure 6a** A ($n_c = 8$) cluster partition of the phase space of the logistic map ($r = 4.0$, N = 2048, $d = 2$, $t = 1$). The normalized eigen-values for obtained by global SVD are $l_1^* \sim$ 0.51 and $l_2^* \sim$ 0.49 and those obtained by local SVD of a cluster are $l_1^{6*} = 0.86$ and $l_2^{6*} =$ 0.14. Adjacent clusters are shown by thin and thick lines in that order.

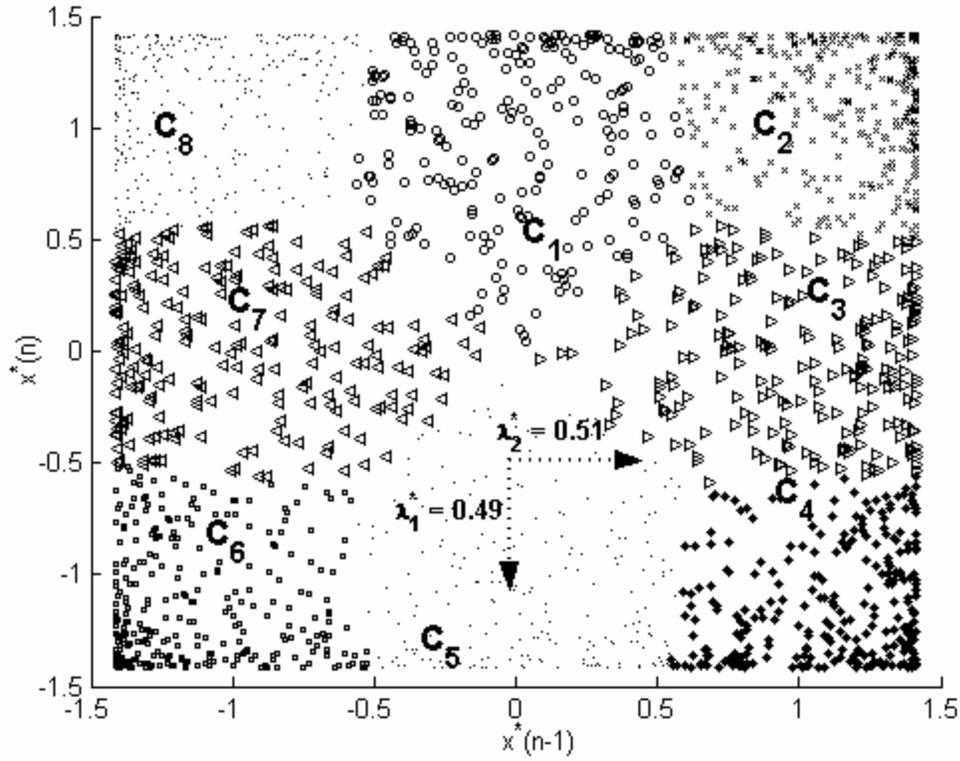

**Figure 6b** A ($n_c = 8$) cluster partition of the phase space of the IAAFT surrogate of the logistic map ($r = 4.0$, N = 2048, $d = 2$, $t = 1$). The normalized eigen-values obtained by global SVD are $l_1^* \sim 0.51$ and $l_2^* \sim 0.49$. The clusters are shown by distinct symbols.

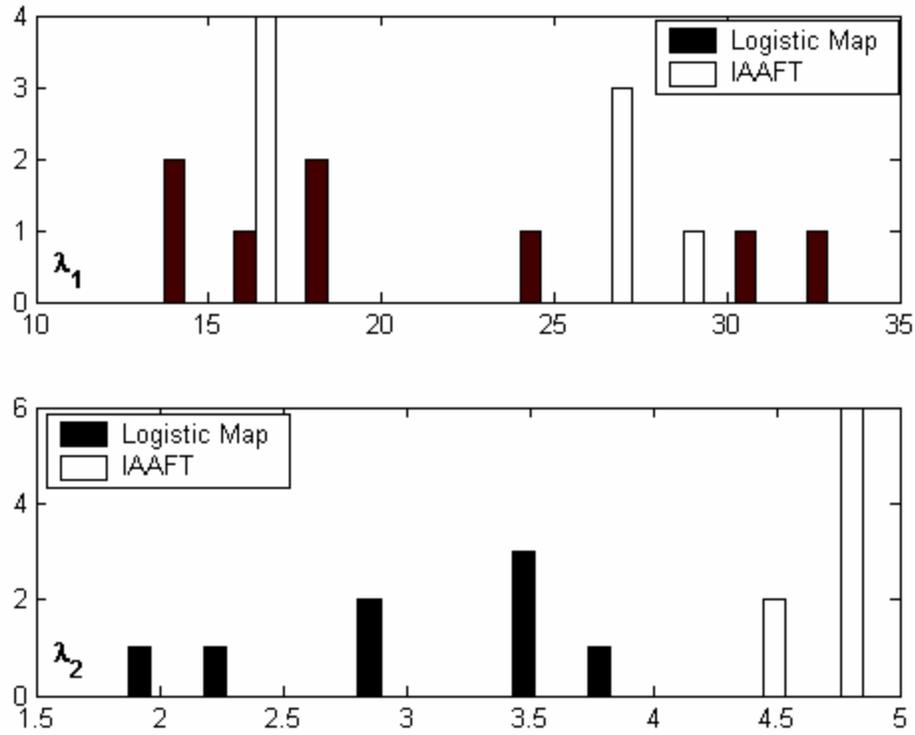

**Figure 6c** The distribution of the eigen-values $\boldsymbol{l}_1^i, i = 1...n_c$ (top) and $\boldsymbol{l}_2^i, i = 1...n_c$

(bottom) for a $(n_c = 8)$ cluster partition of the chaotic logistic map (black bar) and its

IAAFT surrogate (white bar), with embedding parameters (N = 2048, $d = 2$, $\boldsymbol{t} = 1$).

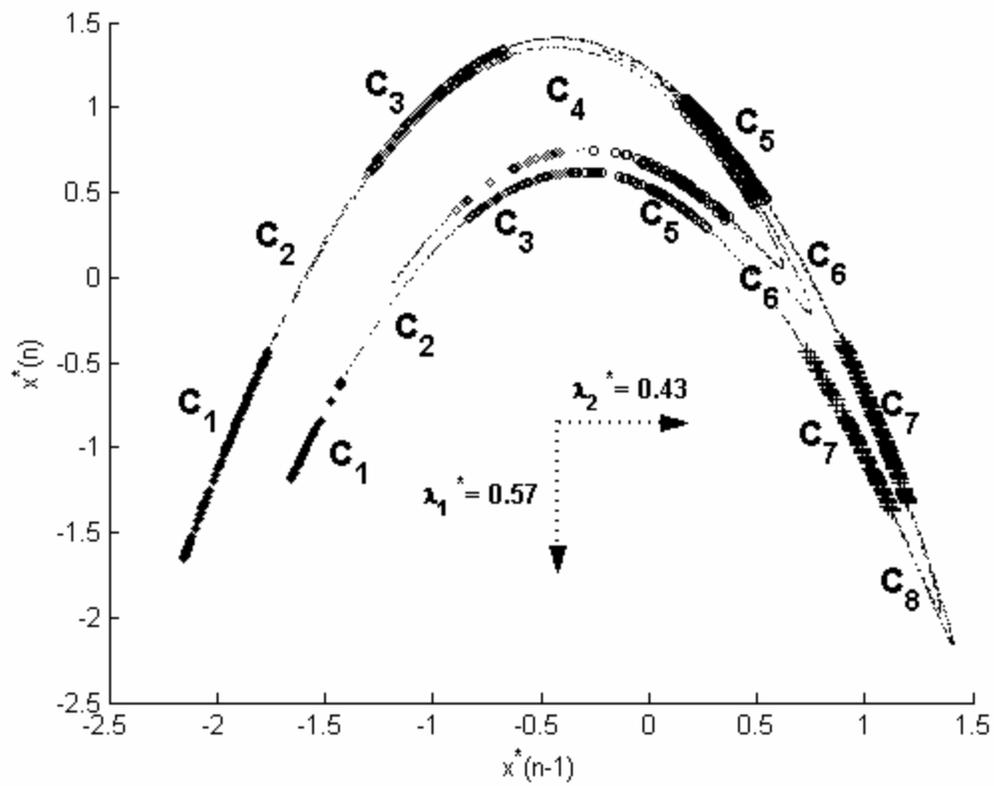

**Figure 7a** A ($n_c = 8$) cluster partition of the phase space of the henon map ($a$=1.4, $b$=0.3, N = 2048, $d$ = 2, $t$ = 1). The normalized eigen-values for obtained by global SVD are $l_1^*$ ~ 0.57 and $l_2^*$ ~ 0.43.

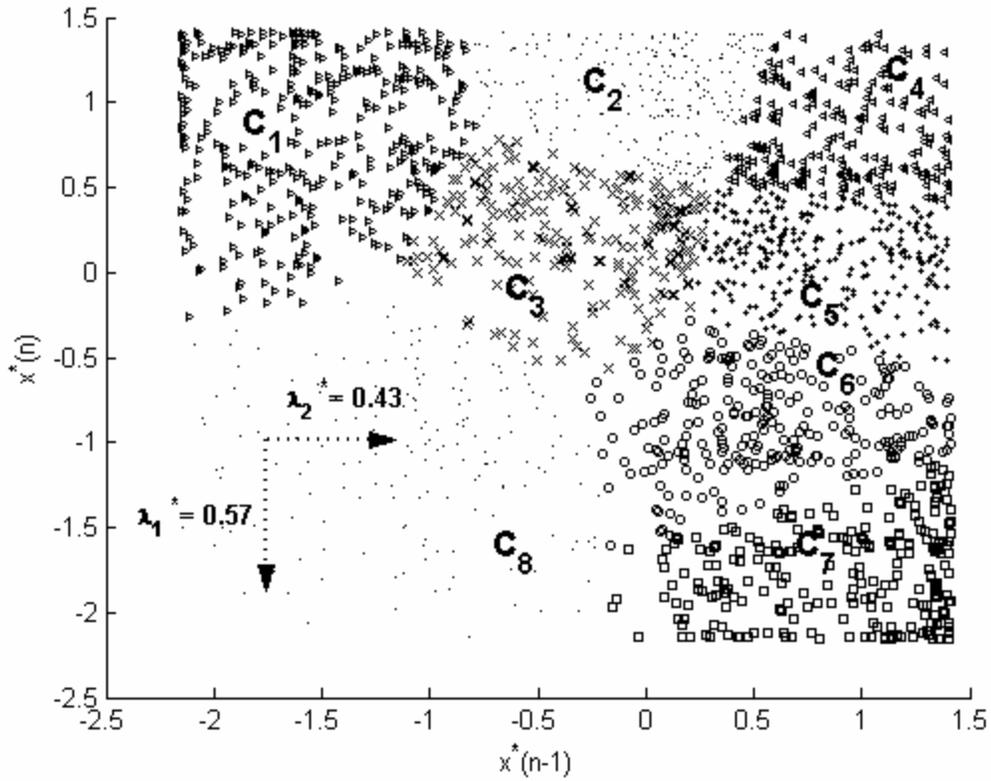

**Figure 7b** A ($n_c = 8$) cluster partition of the phase space of the IAAFT surrogate of the henon map ($a$=1.4, $b$=0.3, N = 2048, $d$ = 2, $t$ = 1). The normalized eigen-values for obtained by global SVD are $\lambda_1^* \sim 0.57$ and $\lambda_2^* \sim 0.43$.

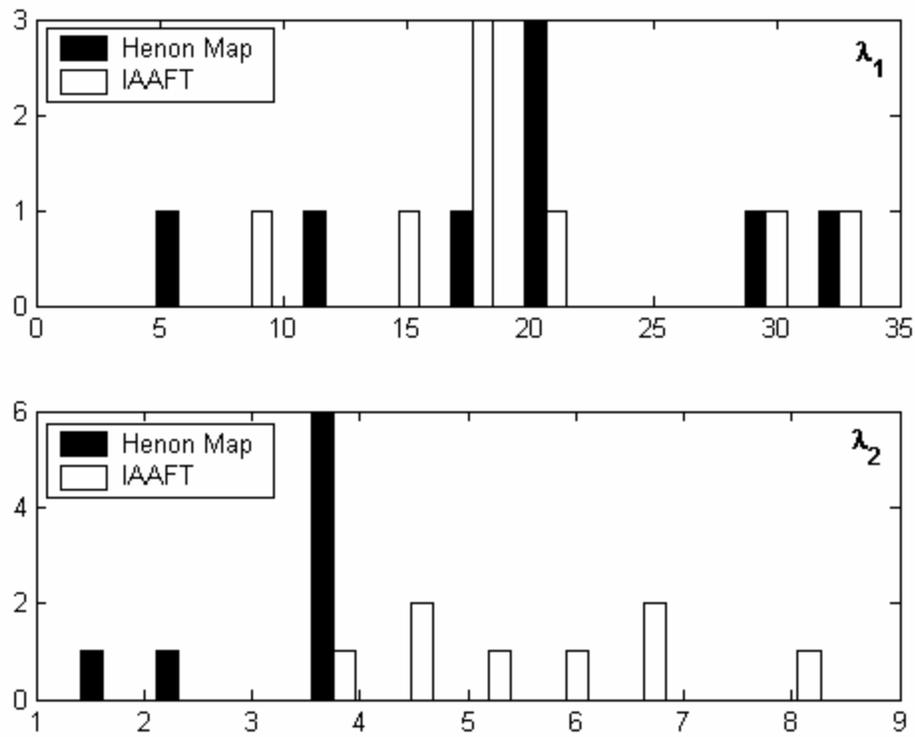

**Figure 7c** The distribution of the eigen-values $l_1^i, i=1...n_c$ (top) and $l_2^i, i=1...n_c$

(bottom) for a ($n_c = 8$) cluster partition of the chaotic logistic map (black bar) and its

IAAFT surrogate (white bar), with embedding parameters (N = 2048, $d = 2$, $t = 1$).

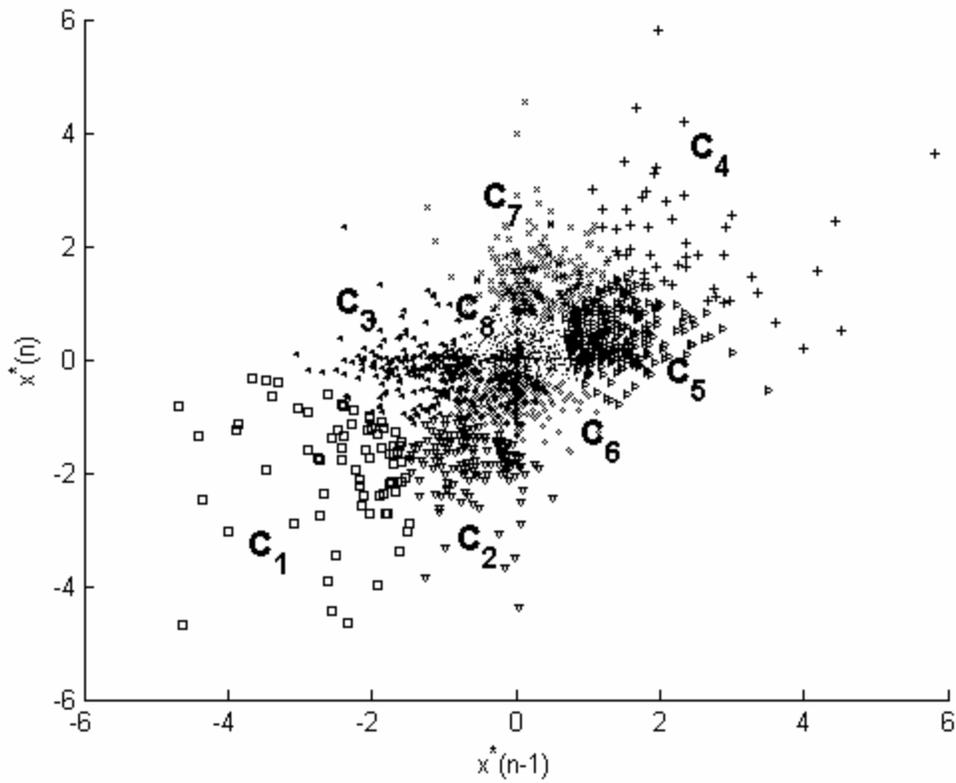

**Figure 8a** A ($n_c = 8$) cluster partition of the phase space of the nonlinearly correlated noise (N = 2048, $d = 2$, $t = 1$). The normalized eigen-values for obtained by global SVD are $l_1^* \sim 0.63$ and $l_2^* \sim 0.37$.

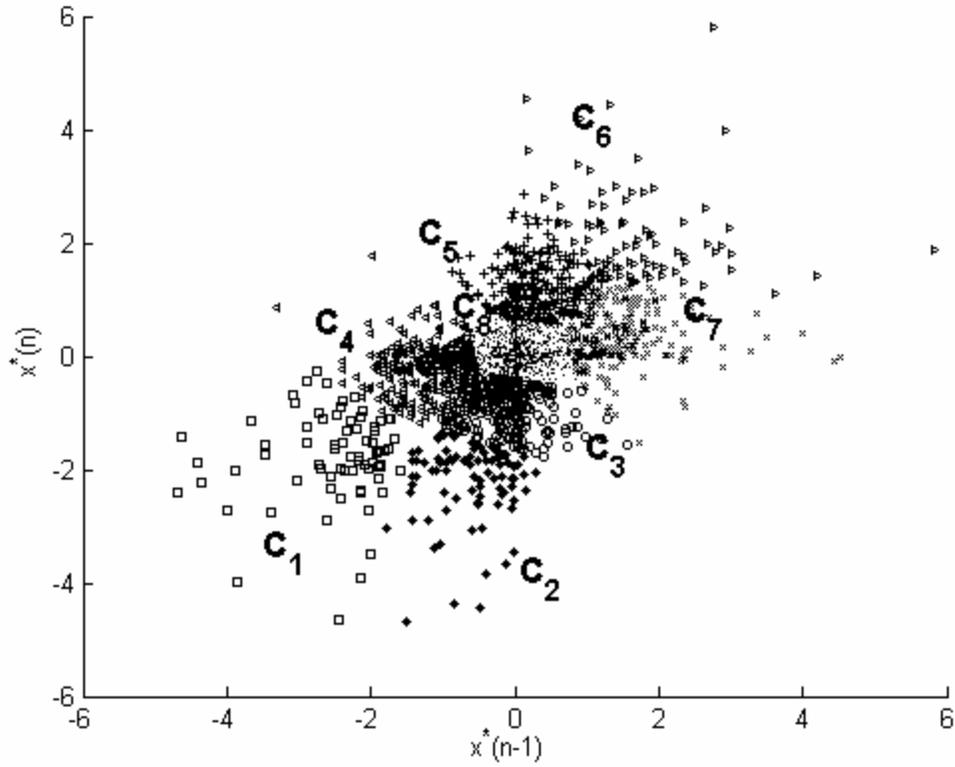

**Figure 8b** A ($n_c = 8$) cluster partition of the phase space of an IAAFT surrogate of the nonlinearly correlated noise (N = 2048, $d = 2$, $t = 1$). The normalized eigen-values for obtained by global SVD are $l_1^* \sim 0.63$ and $l_2^* \sim 0.37$.

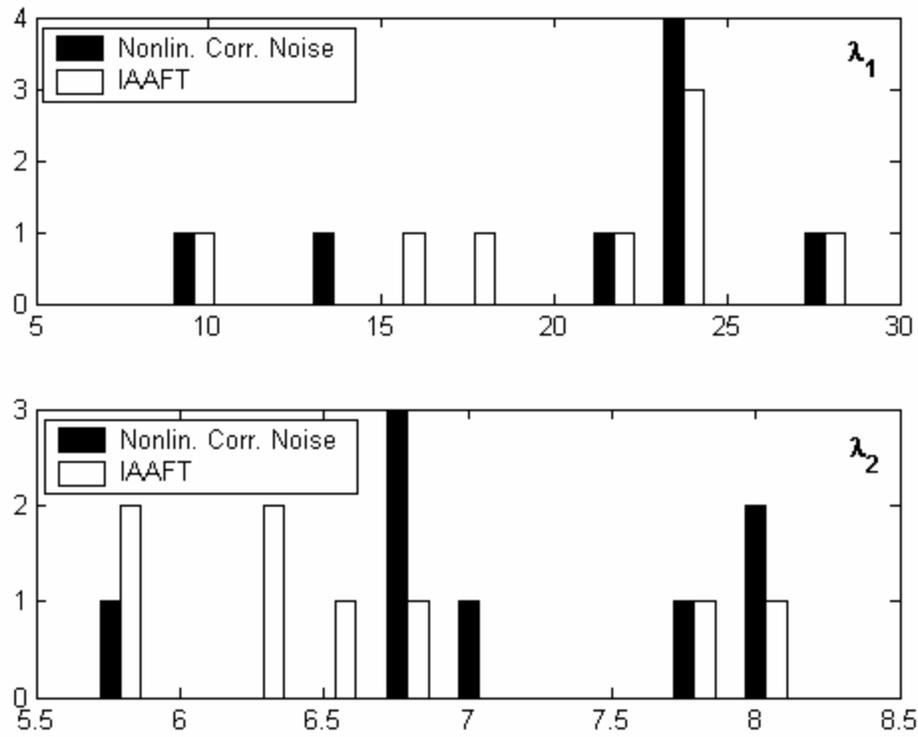

**Figure 8c** The distribution of the eigen-values $l^i_1, i = 1...n_c$ (top) and $l^i_2, i = 1...n_c$ (bottom) for a ($n_c = 8$) cluster partition of the nonlinearly correlated noise (black bar) and its IAAFT surrogate (white bar), with embedding parameters (N = 2048, $d = 2$, $t = 1$).

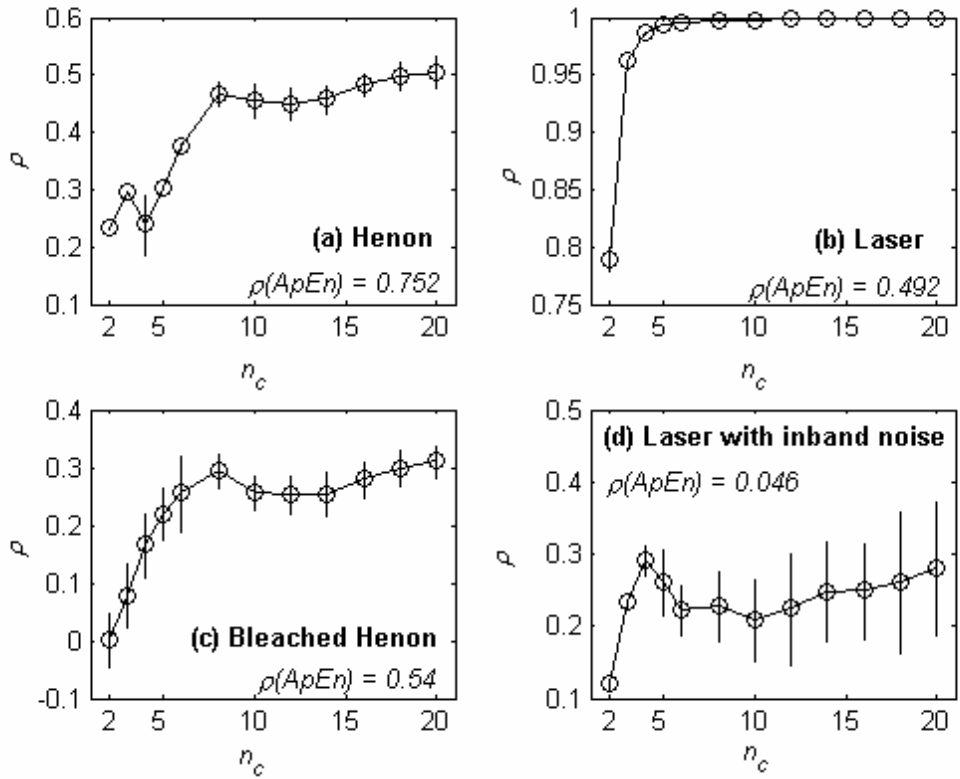

**Figure 9** Variation of ***r*** with ($n_c$ = 2, 3, 4, 5, 6, 8, 10, 12, 14, 16, 18, 20) for the data sets generated by nonlinear deterministic chaotic processes. The vertical lines denoted the standard deviation about the mean (circle) for 100 realizations, obtained using $\boldsymbol{D}^T$ as the discriminant statistic. The ***r*** values obtained on chaotic henon map, chaotic laser, bleached chaotic henon map and chaotic laser with inband noise are shown in (a, b, c and d) respectively. The ***r*** values obtained using approximate entropy as the discriminant statistic is also shown for reference.

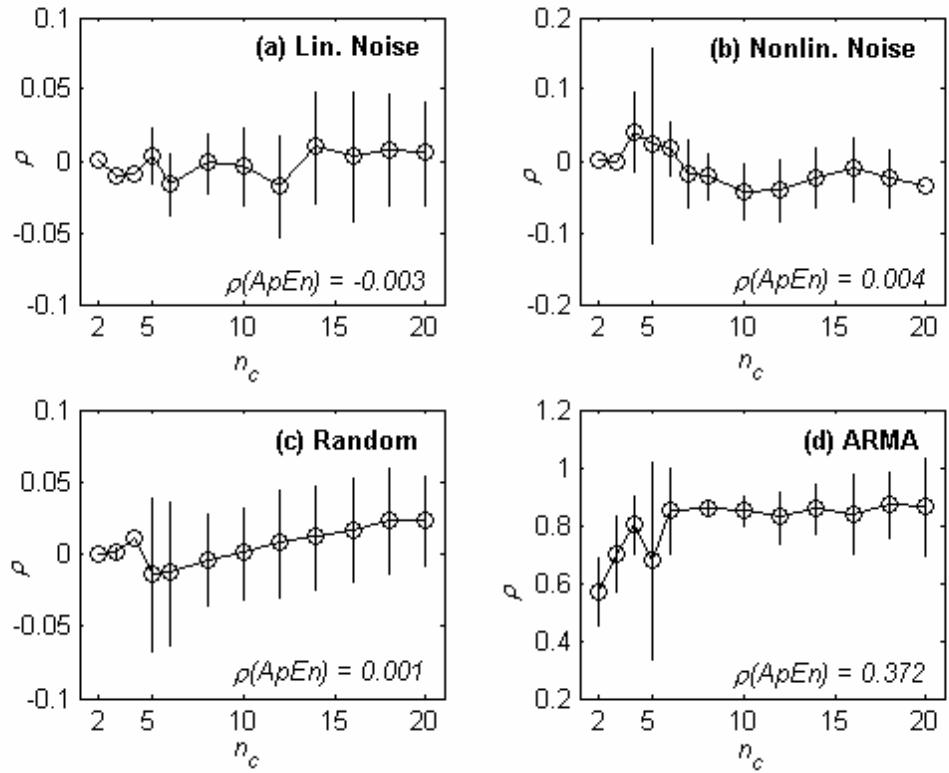

**Figure 10** Variation of **r** with ($n_c$ = 2, 3, 4, 5, 6, 8, 10, 12, 14, 16, 18, 20) for the data sets generated by non-determinisitic processes. The vertical lines denoted the standard deviation about the mean (circle) for 100 realizations, obtained using $\boldsymbol{D}^T$ as the discriminant statistic. The **r** values obtained on linear correlated noise, nonlinear correlated noise, random noise and narrow band ARMA process are shown in (a, b, c and d) respectively. The **r** values obtained using approximate entropy as the discriminant statistic is also shown for reference.

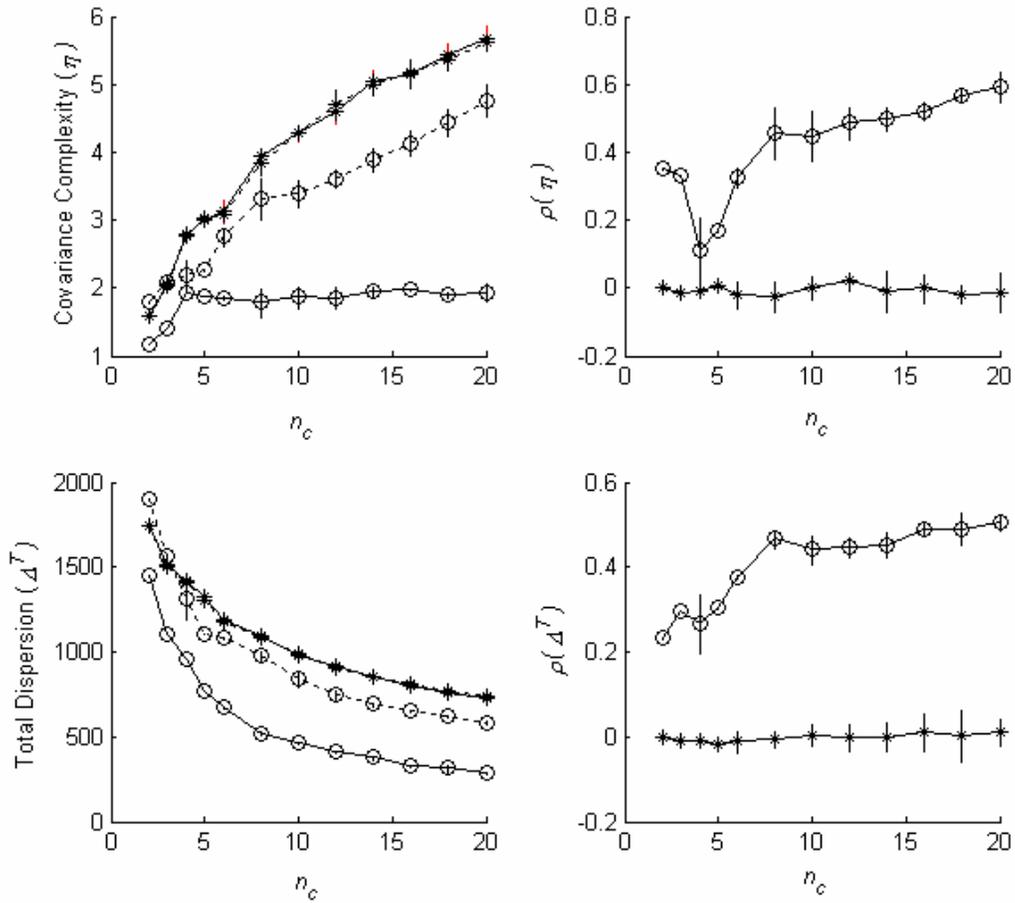

**Figure 11** Variation of Morgera's covariance complexity ($\boldsymbol{h}$) and the total dispersion ($\boldsymbol{D}^T$) with ($n_c$ = 2, 3, 4, 5, 6, 8, 10, 12, 14, 16, 18, 20) for the original data (solid line) and its IAAFT surrogate (dotted line) (left). The corresponding variation in the index $\boldsymbol{r}$ obtained using $\boldsymbol{h}$ and $\boldsymbol{D}^T$ as discriminant statistic are shown on the right. The vertical lines represent the mean value obtained on 10 different realizations. The mean value obtained on chaotic henon map ($d$ = 2, $\boldsymbol{t}$ = 1, N = 2048) is represented by (o) and those obtained on linearly correlated noise ($d$ = 2, $\boldsymbol{t}$ = 1, N = 2048) is represented by (*).

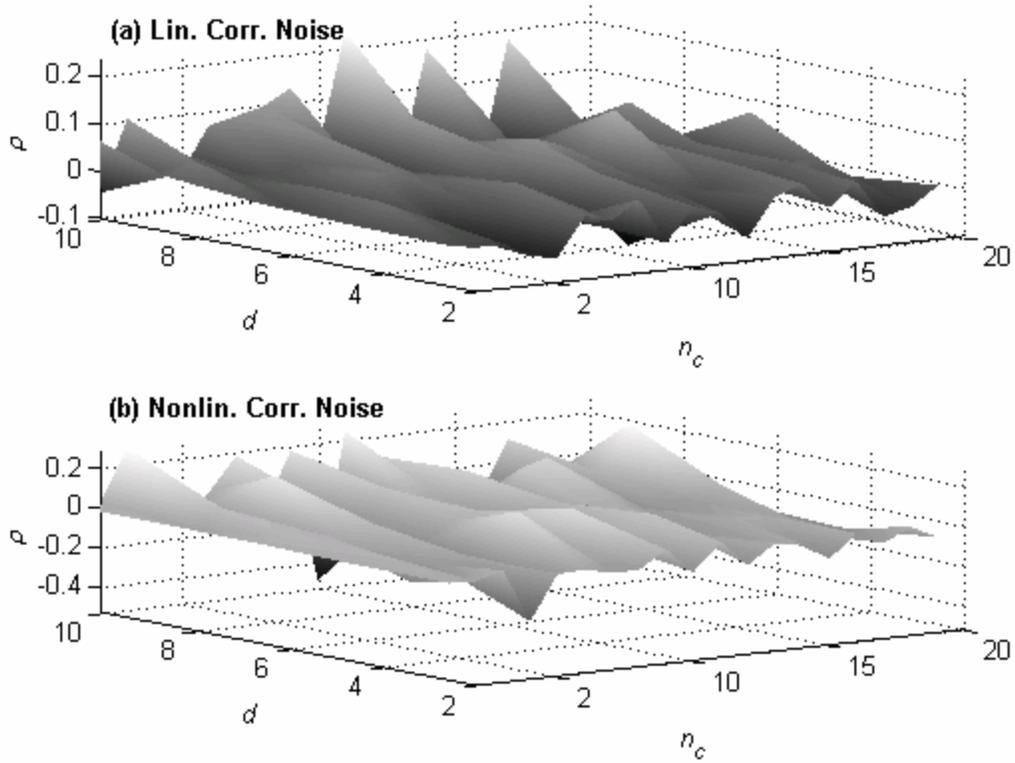

**Figure 12** Variation in **r** with varying embedding dimension ($d$ = 2, 4, 6, 8 and 10, N = 2048, **t** = 1) and varying number of clusters ($n_c$ = 2…20) for the linear correlated noise (a) and the nonlinear correlated noise (b). The average value on 10 realizations obtained using the total dispersion ($\mathbf{D}^T$) as the discriminant statistic is shown.

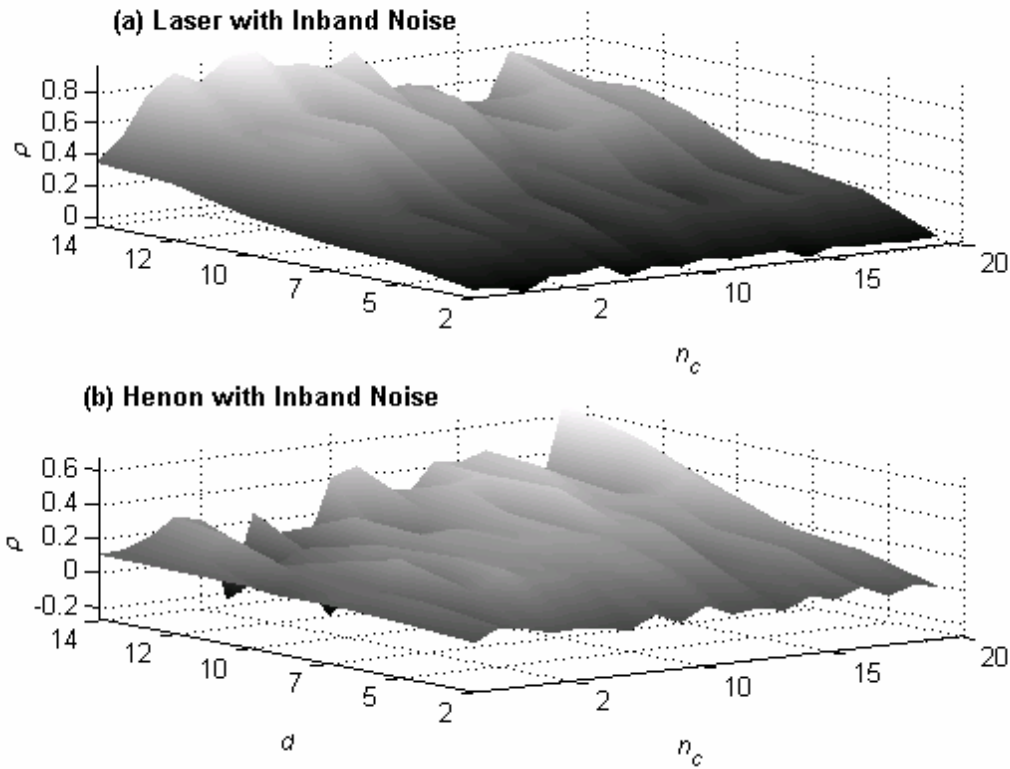

**Figure 13** Variation in ***r*** with varying embedding dimension (*d* = 2, 5, 7, 10, 12 and 14, N = 2048, ***t*** = 1) and varying number of clusters (*n_c* = 2…20) for the chaotic laser with inband noise (a) and chaotic henon with inband noise (b). The average value on 10 realizations obtained using the total dispersion (***D^T***) as the discriminant statistic is shown.

**Table I**

**Statistical comparison of distribution of the eigen-values between the original data and its IAAFT surrogate for a ($n_c = 8$) partition. $l^M$ and $l^L$ represent the most dominant and the least dominant eigen-values.**

| Data | $l$ | Ttest | wilcoxon | Resamp | Null |
|------|-----|-------|----------|--------|------|
| **Chaotic Logistic** | $l^M$ | 0.47 | 0.94 | 0.47 | Cannot Reject |
| $m = 2$, $t = 1$ | $l^L$ | $p < 10^{-2}$ | $p < 10^{-2}$ | $p < 10^{-2}$ | Reject |
| **Chaotic Henon** | $l^M$ | $p = 0.47$ | $p = 0.87$ | $p = 0.46$ | Cannot Reject |
| $m = 2$, $t = 1$ | $l^L$ | $p < 10^{-2}$ | $p < 10^{-2}$ | $p < 10^{-2}$ | Reject |
| **Chaotic Laser** | $l^M$ | $p = 0.71$ | $p = 0.64$ | $p = 0.66$ | Cannot Reject |
| $m = 7$, $t = 1$ | $l^L$ | $p < 10^{-2}$ | $p < 10^{-2}$ | $p < 10^{-2}$ | Reject |
| **Bleach. Chao. Henon** | $l^M$ | $p = 0.62$ | $p = 0.95$ | $p = 0.61$ | Cannot Reject |
| $m = 2$, $t = 1$ | $l^L$ | $p < 10^{-2}$ | $p < 10^{-2}$ | $p < 10^{-2}$ | Reject |
| **Laser Inband Noise** | $l^M$ | $p = 0.49$ | $p = 0.88$ | $p = 0.49$ | Cannot Reject |
| $m = 7$, $t = 1$ | $l^L$ | $p = 0.12$ | $p = 0.23$ | $p = 0.12$ | Cannot Reject |
| **Lin. Corr. Noise** | $l^M$ | $p = 0.46$ | $p = 0.95$ | $p = 0.46$ | Cannot Reject |
| $m = 2$, $t = 1$ | $l^L$ | $p = 0.62$ | $p = 0.95$ | $p = 0.62$ | Cannot Reject |
| **Non.Corr.Noise** | $l^M$ | $p = 0.47$ | $p = 0.96$ | $p = 0.47$ | Cannot Reject |
| $m = 2$, $t = 1$ | $l^L$ | $p = 0.87$ | $p = 0.20$ | $p = 0.87$ | Cannot Reject |
| **White Noise** | $l^M$ | $p = 0.48$ | $p = 0.88$ | $p = 0.47$ | Cannot Reject |
| $m = 2$, $t = 1$ | $l^L$ | $p = 0.52$ | $p = 0.23$ | $p = 0.52$ | Cannot Reject |

**Table II**

**Statistical comparison of distributions of the local dispersions obtained on the original and the surrogate realizations for ($n_c = 20$) partition**

| Data | m | $t$ | ttest | wilcoxon | resamp | Null |
|---|---|---|---|---|---|---|
| **Chaotic Logistic** | *2* | *1* | $p < 10^{-2}$ | $p < 10^{-2}$ | $p < 10^{-2}$ | Reject |
| **Chaotic Henon** | *2* | *1* | $p < 10^{-2}$ | $p < 10^{-2}$ | $p < 10^{-2}$ | Reject |
| **Chaotic Laser** | *7* | *1* | $p < 10^{-2}$ | $p < 10^{-2}$ | $p < 10^{-2}$ | Reject |
| **Bleach. Chao. Henon** | *2* | *1* | $p = 0.02$ | $p = 0.05$ | $p = 0.005$ | Cannot Reject |
| **Laser Inband Noise** | *7* | *1* | $p = 0.08$ | $p = 0.08$ | $p = 0.04$ | Cannot Reject |
| **Lin. Corr. Noise** | *2* | *1* | $p = 0.97$ | $p = 0.78$ | $p = 0.47$ | Cannot Reject |
| **Non.Corr.Noise** | *2* | *1* | $p = 0.84$ | $p = 0.74$ | $p = 0.44$ | Cannot Reject |
| **White Noise** | *2* | *1* | $p = 0.95$ | $p = 0.72$ | $p = 0.50$ | Cannot Reject |

**Table III**

**Statistical comparison of the total dispersion obtained on the original data to those obtained on an ensemble of surrogate realizations ($n_c = 20$ and $a = 0.01$)**

| Data | m | $t$ | Parametric (S) | Non-Parametric |
|---|---|---|---|---|
| **Chaotic Logistic** | 2 | 1 | 19.4 | Reject |
| **Chaotic Henon** | 2 | 1 | 15.4 | Reject |
| **Chaotic Laser** | 7 | 1 | 6.1 | Reject |
| **Bleach. Chao. Henon** | 2 | 1 | 11.9 | Reject |
| **Laser Inband Noise** | 7 | 1 | 3.3 | Reject |
| **Lin. Corr. Noise** | 2 | 1 | -1.4 | Cannot Reject |
| **Non.Corr.Noise** | 2 | 1 | -0.18 | Cannot Reject |
| **White Noise** | 2 | 1 | 1.4 | Cannot Reject |

# Appendix

## A. Covariance Matrix and Correlation Matrix

The covariance matrix of the trajectory matrix $\Gamma_{N-(d-1)t \, x \, d}$ (3) is given by $\Gamma_c$ where

$$\boldsymbol{g}_{c(ij)} = \boldsymbol{g}_{(ij)} - \boldsymbol{m}_i, \text{ where } \boldsymbol{m}_i = \sum_{j=1}^{N-(d-1)t} \boldsymbol{g}_{ji}$$. Subtracting the mean shifts the centroid of

attractor to the origin but does not distort its geometry. The correlation matrix $\Gamma_r$,

is defined such that $\boldsymbol{g}_{r(ij)} = \boldsymbol{g}_{c(ij)} / \boldsymbol{s}_i$, where $\boldsymbol{s}_i$ is the auto-correlation of the $i^{th}$ column, $\Gamma_{c,i}$. The correlation matrix has the nice property that the sum of the eigen-values of $\Gamma_r$ to be equal to the dimension $d$. The choice of the correlation matrix is valid provided the variance across each of the dimensions are identical, which is likely to be true for large data sets. In the case of short data sets, using the correlation or the covariance matrices can introduce numerical errors and hence should be avoided.

*Mapping of a unit sphere into an ellipsoid under a linear transformation $\boldsymbol{G}$*

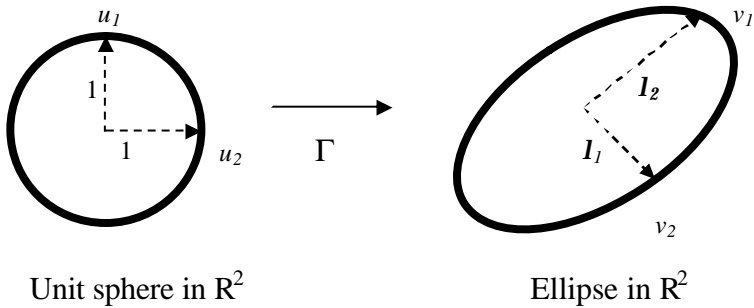

Unit sphere in R$^2$              Ellipse in R$^2$

In general, for $\Gamma \in \text{R}^{Nxd}$, the singular values $\boldsymbol{l}_i, i = 1..d$ of $\Gamma$ represent the length of the axes of a $d$-dimensional ellipsoid, along orthogonal directions $v_i, i = 1..d$.

**B. Wiener–Khintchine Relation**

The power spectrum and the auto-correlation of a wide-sense stationary (WSS) random process is given by the relation

$$P_{xx}(f) = \int_{-\infty}^{\infty} r_{xx}(\boldsymbol{t}) e^{-j2\boldsymbol{p}ft} d\boldsymbol{t} \text{ where } r_{xx}(\boldsymbol{t}) = E[x^*(t)x(t+\boldsymbol{t})]$$

In the above relation, $P_{xx}(f)$ represents the power spectrum and $r_{xx}(\boldsymbol{t})$ the auto-correlation function.

**C. Caratheodory's Uniqueness Result**

Given a one-dimensional time series of the form $x_n = \sum_{k=1}^{p} |c_k|^2 e^{j\boldsymbol{w}_k(n-1)}, n = 1...m$, if

$m > p+1$, then $\boldsymbol{w}_k \in [-\boldsymbol{p}, \boldsymbol{p})$ and $|c_k|^2, k = 1...p$, are unique.

**D. Spectral Estimation**

The Fourier series expansion decomposes a given periodic signal $x(t)$ with period

$p$ and fundamental frequency $\boldsymbol{w}_0 = \dfrac{2\boldsymbol{p}}{p} = 2\boldsymbol{p}f_0$, as a *linear* combination of

orthogonal basis functions, given by the infinite series

$$x(t) = \sum_{k=0}^{\infty} a_k \cos(\boldsymbol{w}_0 kt) + \sum_{k=1}^{\infty} b_k \sin(\boldsymbol{w}_0 kt)$$

An alternate expression can be obtained in terms of complex exponentials $e^{j\boldsymbol{w}_0 kt}$, by invoking the Euler's formula. The discrete sequence, $x(n)$ is obtained by sampling $x(t)$ at a rate greater than twice the maximum frequency component in the signal, i.e. $x(n) = x(t = nT_s)$ where $F_s = 1/T_s$ is the sampling frequency. This is done so as to prevent *aliasing* [Proakis and Manolakis, 1995]. The discrete time periodic signal is given by

$$x(n) = \sum_{k=0}^{\infty} a_k \cos(2\boldsymbol{p}nk\frac{f_0}{F_s}) + b_k \sin(2\boldsymbol{p}nk\frac{f_0}{F_s})$$

The frequencies $k\frac{f_0}{F_s}$ represent the $k^{th}$ harmonic of the fundamental $f_o$, hence the above expression is also known as *harmonic decomposition*.

**D1.** *Subspace Decomposition*

Consider a wide-sense stationary (WSS) process $x(n) = \sum_{i=1}^{p} A_i e^{j(2\boldsymbol{p}f_i n + \boldsymbol{F}_i)}$, with unknown parameters $A_i, f_i$ representing the amplitude and frequency with phases $\boldsymbol{f}_i$ uniformly distributed in $[0, 2\boldsymbol{p}]$, respectively.

An *m*-dimensional reconstruction with $(m > p)$ is given by

$$\mathbf{x} = [x(n) \, x(n-1) .... x(n-1+m)]^T$$

The *mxm* lagged correlation matrix $R_{xx}$ has a toeplitz structure and is given by

$$R_{xx} = E(\mathbf{x}\mathbf{x}^H)$$

where $H$ represents conjugate transpose of $\mathbf{x}$ and E the expectation operator. In the presence of white noise, define $y_n = x_n + \in_n$, where $\in_n$ is zero mean additive white noise with auto-correlation $\boldsymbol{s}_e^2$. The lagged correlation matrix of the noisy data can be represented as the additive sum $R_{yy} = R_{xx} + \boldsymbol{s}_e^2 I$.

The signal matrix, $R_{xx} = \sum_{i=1}^{p} P_i s_i^H s_i$, where $s_i$ is an $m$-dimensional signal vector given by $s_i = [1, e^{j2\boldsymbol{p}f_i}, \ldots, e^{j2\boldsymbol{p}(m-1)f_i}]$. The $p$ *dominant* eigen-vectors span the *signal subspace*, the remaining $(m\text{-}p)$ eigen-vectors span the *noise subspace*. For real valued data, $R_{xx}$ is the symmetric matrix $\Gamma^T \Gamma$ (3), with ($t=1$, $d = m$). Estimating the eigen-values by SVD of $\Gamma$ is robust compared to eigen-decomposition of $R_{xx}$.

**D2.** *Representing real sinusoidal data with additive white noise as an ARMA(p,p) process*

A real signal consisting of $p$ sinusoidal components can be represented by the difference equation

$$x(n) = -\sum_{k=1}^{2p} a_k x(n-k)$$

A sine wave with additive noise, is given by $y(n) = x(n) + e(n)$, where $e(n)$ is zero mean additive white noise with variance $\boldsymbol{s}_e^2$.

Substituting $x(n) = y(n) - e(n)$ yields

$$\sum_{k=0}^{2p} a_k x(n-k) = \sum_{l=0}^{2p} a_k e(n-l), \text{ with } a_0 = 1.$$

The above expression represents a auto-regressive moving average process *ARMA(p,p)*, with identical auto-regressive (*AR*) and moving-average (*MA*) parameters.

In general an ARMA process in general is given by the expression

$$x(n) = \sum_{k=1}^{M} a_k x(n-k) + \sum_{l=0}^{N} b_k e(n-l)$$

## E. Covariance complexity

A brief description of the covariance complexity [Morgera, 1985; Watanabe et al., 2003] is enclosed below. Let the local trajectory matrix on a $n_c$ partition of a $m$-dimensional phase space be $\Gamma^k, k=1...n_c$. SVD of $\Gamma_i$ yields eigen-values $l_i^k, i=1...m, k=1...n_c$. The variance along the $i^{th}$ component in the $n^{th}$ cluster is given by

$$s_i^n = \frac{(l_i^n)^2}{\sum_{i=1}^{m}(l_i^n)^2}, i=1...m$$

Morgera's covariance complexity for the $n^{th}$ cluster is given by

$$C^n = -\frac{1}{\log m} \sum_{p=1}^{m} s_p^n \log s_p^n$$

The covariance complexity of the entire phase space is given by

$$h = \sum_{n=1}^{n_c} C^n$$

## F. Synthetic Data Sets

*Logistic Map*      $x_{n+1} = r x_n (1 - x_n)$                    (F1)

*Henon Map*     [Henon, 1976]                    (F2)

$$x_{n+1} = 1 - a x_n^2 + y_n$$
$$y_{n+1} = b x_n$$

*Periodic Sine:* $x_n = \sin(\frac{3\boldsymbol{p}}{50}n) + \sin(\frac{5\boldsymbol{p}}{50}n) + \sin(\frac{8\boldsymbol{p}}{50}n) + \sin(\frac{11\boldsymbol{p}}{50}n)$       (F3)

*Quasiperiodic Sine:* $x_n = \sin(\frac{\sqrt{3}\boldsymbol{p}}{50}n) + \sin(\frac{5\boldsymbol{p}}{50}n) + \sin(\frac{8\boldsymbol{p}}{50}n) + \sin(\frac{11\boldsymbol{p}}{50}n)$   (F4)

*ARMA (Sinusoid corrupted with noise)*                                 (F5)

$$x_n = -x_{n-1} - x_{n-2} - x_{n-3} - x_{n-4} + \boldsymbol{a} \in_n$$

where $\in_n$ is zero mean unit variance additive white gaussian noise, with noise factor $\boldsymbol{a} = 0.7$ and initial conditions $x_{-1} = 0.5, x_{-2} = -0.2, x_{-3} = 0.5, x_{-2} = -0.2$.

*Linearly correlated noise*                                       (F6)

$$x_n = 0.5 x_{n-1} + \in_n$$

where $\in_n$ is zero mean unit variance additive white gaussian noise.

*Nonlinear transform of correlated noise* [Schreiber & Schmitz, 1996]      (F7)

$$x_n = 0.5 x_{n-1} + \in_n, \ y_n = x_n \sqrt{|x_n|}$$

where $\in_n$ is zero mean unit variance additive white gaussian noise.

*Laser Data* [Huebner et al, 1989]                                (F8)

The data set was recorded from a Far-Infrared-Laser in a chaotic state. The

measurements were made on an 81.5-micron 14NH3 cw (FIR) laser, pumped

optically by the P(13) line of an N2O laser via the vibrational a Q(8,7) NH3

transition. The experimental signal to noise ratio was about 300, slightly under the half bit uncertainty of the analog to digital conversion. Chaotic pulsations more or less follow the theoretical Lorenz model.

*Bleached Data* [Theiler and Eubank, 1993]                                    (F9)

To test the effect of bleaching, a linear predictor of optimal order six ($y_p$), was fit to the chaotic henon data (y). The predicted values were subtracted from the original data to obtain the bleached data, i.e. $y_{bleach} = y - y_p$.

*Inband Noise* [Schreiber & Schmitz, 1997]                                    (F10)

The in-band noise ($\in_n$) is a phase-randomized copy of the given data ($y_n$). The noisy data, with noise factor (*a*) is generated by the expression

$$x_n(a) = \sqrt{\frac{1}{1+a^2}}\,(y_n + a\in_n)$$

The value of the noise factor was fixed at (*a* = 0.8) for the laser data (F7) with inband noise and (*a* = 1.0) for the henon data (F2) with inband noise.